\documentclass[12pt]{article}
\usepackage{amsmath,amssymb}
\usepackage{graphicx,epsfig}
\usepackage{axodraw}
\def\simgt{\mathrel{\lower2.5pt\vbox{\lineskip=0pt\baselineskip=0pt
           \hbox{$>$}\hbox{$\sim$}}}}
\def\simlt{\mathrel{\lower2.5pt\vbox{\lineskip=0pt\baselineskip=0pt
           \hbox{$<$}\hbox{$\sim$}}}}

\begin{document}
\baselineskip 0.6cm

\begin{titlepage}

\begin{flushright}
UCB-PTH-03/04 \\
LBNL-52212 \\
FERMILAB-Pub-03/033-T
\end{flushright}

\vskip 1.0cm

\begin{center}
{\Large \bf Explicit Supersymmetry Breaking\\
 on Boundaries of Warped Extra Dimensions}

\vskip 1.0cm

{\large
Lawrence J.~Hall$^*$, 
Yasunori Nomura$^\dagger$, 
Takemichi Okui$^*$, 
and Steven J.~Oliver$^*$}

\vskip 0.5cm

$^*$ {\it Department of Physics, University of California, Berkeley,\\
  and\\
    Theoretical Physics Group, Lawrence Berkeley National Laboratory,\\
    Berkeley, CA 94720, USA}\\
$^\dagger$ {\it Theoretical Physics Department, 
    Fermi National Accelerator Laboratory,\\
    Batavia, IL 60510, USA}

\vskip 1.0cm

\abstract{Explicit supersymmetry breaking is studied in higher 
dimensional theories by having boundaries respect only a subgroup 
of the bulk symmetry. If the boundary symmetry is the maximal 
subgroup allowed by the boundary conditions imposed on the fields, 
then the symmetry can be consistently gauged; otherwise gauging 
leads to an inconsistent theory. In a warped fifth dimension, 
an explicit breaking of all bulk supersymmetries by the boundaries 
is found to be inconsistent with gauging; unlike the case of flat 5D, 
complete supersymmetry breaking by boundary conditions is not consistent 
with supergravity. Despite this result, the low energy effective theory
resulting from boundary supersymmetry breaking becomes consistent 
in the limit where gravity decouples, and such models are explored 
in the hope that some way of successfully incorporating gravity 
can be found. A warped constrained standard model leads to a theory 
with one Higgs boson with mass expected close to the experimental 
limit. A unified theory in a warped fifth dimension is studied with 
boundary breaking of both $SU(5)$ gauge symmetry and supersymmetry. 
The usual supersymmetric prediction for gauge coupling unification 
holds even though the TeV spectrum is quite unlike the MSSM. 
Such a theory may unify matter and Higgs in the same $SU(5)$ 
hypermultiplet.}

\end{center}
\end{titlepage}

\section{Introduction}
\label{sec:intro}

A light Higgs boson, suggested by precision electroweak data, together
with a heavy top quark, has direct and consequential implications. 
Virtual top quarks necessarily induce a large quadratic divergence to
the Higgs mass parameter, hence a light Higgs boson is expected only
if this is canceled by additional radiative contributions from new 
physics at energy scales not far above the top quark mass. The most
obvious origin for this cancellation is weak scale supersymmetry, and
the case for this is greatly strengthened by the successful prediction 
from gauge coupling unification.  The experimental implications for 
such theories have focused almost exclusively on theories which are 
four dimensional at the TeV scale --- especially the minimal 
supersymmetric standard model (MSSM). However, higher dimensional 
supersymmetric theories can also tame the divergences of scalar 
mass parameters, with cancellations from Kaluza-Klein (KK) 
modes playing as important a role as cancellations from 
superpartners~\cite{Antoniadis:1998sd}.  Furthermore, beneath the 
mass scale of the lightest KK modes, the 4D effective theory need 
not be supersymmetric --- there is no MSSM limit of the theory.

In the MSSM the weak scale is understood as a byproduct of the more
fundamental supersymmetry breaking scale. When KK modes play a crucial
role in canceling the Higgs mass divergence, the more fundamental scale 
is the effective compactification scale, $M_c$, which is the mass threshold 
for the KK modes. This mass scale, which should not be far above the 
top quark mass, should trigger the breaking of both supersymmetry 
and electroweak symmetry. In the constrained standard model of 
Ref.~\cite{Barbieri:2000vh} $M_c \simeq 350~{\rm GeV}$, and it is obvious 
that there is no MSSM limit: there is only one Higgs doublet, and it 
couples to both up and down type quarks. Two Higgs theories can also be 
constructed~\cite{Arkani-Hamed:2001mi}, as can theories with $M_c$ 
considerably higher, in the $30~{\rm TeV}$ region~\cite{Pomarol:1998sd}. 
These latter theories may mimic the MSSM at future collider experiments.

A common feature of these theories is that supersymmetry breaking
arises because the boundary conditions in the fifth dimension are
taken to differ for fermions and bosons. The non-locality of this 
breaking implies supersymmetry breaking counterterms cannot be
induced by radiative corrections. The cancellations in the Higgs mass 
parameters are more precise than in 4D theories, so that the Higgs 
mass parameters are finite and calculable. In the constrained 
standard model there is only a single Higgs field, with the potential 
calculated in terms of a single free parameter $M_c$, making it 
possible to determine $M_c$ by the $Z$ mass and predict the physical 
Higgs boson mass: $127 \pm 8~{\rm GeV}$. 

Despite these successes, one must admit a significant drawback. 
In all these theories the gauge and Yukawa couplings become strongly 
coupled in the multi-TeV domain and the UV cutoff of the effective 
5D field theory is reached long before the unification scale, so that 
the successful prediction from conventional logarithmic unification 
is lost. Furthermore, since the cutoff of the theory is in the 
multi-TeV domain, one must address the question of why gravity is
so weak. This apparently requires further extra dimensions, either 
in the sub mm domain~\cite{Arkani-Hamed:1998rs} or with a warp 
factor~\cite{Randall:1999ee}.

There is a very simple way to maintain gauge coupling unification 
even when supersymmetry is broken by boundary conditions in a fifth
dimension. It is possible that the difference in the boundary
conditions between fermions and bosons is described by a very small
angle $\alpha$~\cite{Barbieri:2001yz}, so that the scale of the 
superpartners, $\alpha M_c$, can become decoupled from the 
compactification scale. Gauge coupling unification is recovered if 
$M_c$ is taken at or above the unification scale. However, in this 
case, since the KK modes are at or above the unification scale, the 
cancellation of the top divergence in the Higgs mass parameter reduces 
precisely to the usual 4D supersymmetric case. In this paper we want 
to ask a different question: is it possible for the KK modes to take 
part in the cancellation of the Higgs mass divergence, while allowing 
conventional logarithmic gauge coupling unification? Furthermore, 
how would the weakness of gravity be understood in such a theory 
with TeV scale KK modes? One possibility is to also have sub mm scale 
extra dimensions for the propagation of gravity, but then it is not 
clear how to recover gauge coupling unification. A second possibility 
is that there is a warped extra dimension, in which case the running 
of gauge couplings is logarithmic above the mass threshold for the KK 
towers~\cite[\ref{Randall:2001gc:X}~--~\ref{Choi:2002ps:X}]{Pomarol:2000hp}.
In general $SU(3)_C \times SU(2)_L \times U(1)_Y$ theories the low 
energy gauge couplings cannot be predicted, because they depend on the 
tree-level 5D gauge couplings, which are free parameters of the 
theory.  However, if the bulk of this warped dimension has unified 
gauge symmetry such as $SU(5)$, one can show that the successful 
prediction of the MSSM for gauge coupling unification can be 
obtained~\cite{Goldberger:2002pc}.  In fact, such theories can be 
constructed by breaking the unified gauge symmetry either by 
the vacuum expectation value of a Planck-brane localized 
field~\cite{Pomarol:2000hp} or by boundary conditions imposed 
at the Planck brane~\cite{Goldberger:2002pc}.  This offers the 
possibility of exceptional economy: the warped dimension that 
generates the TeV scale and the dimension which contains supersymmetry 
breaking boundary conditions could be one and the same.

With the above motivation, in this paper we study boundary condition 
breaking of supersymmetry in warped space, in particular in a slice 
of AdS$_5$. Our aim is to construct a theory of electroweak symmetry 
breaking where a crucial role is played by the TeV mass KK modes of 
this warped extra dimension, while simultaneously solving the gauge 
hierarchy problem and addressing logarithmic gauge coupling unification.
However, before attempting to construct a model, we must study whether 
it is consistent to impose supersymmetry breaking boundary conditions 
in a supersymmetric theory in a warped 5D spacetime.

We note that it is straightforward to construct 5D warped, supersymmetric 
theories with supersymmetry broken spontaneously by a vacuum expectation 
value (VEV) located on the TeV brane~\cite{Gherghetta:2000qt}. With 
gauge interactions in the bulk, but matter and Higgs fields on the 
Planck brane, supersymmetry breaking is mediated to matter and Higgs 
via gaugino mass terms. By introducing the bulk $SU(5)$, 
one can also recover the MSSM prediction for gauge coupling 
unification~\cite{Goldberger:2002pc}. These theories are rather 
interesting, since the 5D nature for the Higgs mass cancellation 
is obtained by taking the VEV to be large.  However, in these theories 
the scale of supersymmetry breaking is in principle a free parameter 
and is not strictly related to the KK mass scale. It is also difficult 
to construct one Higgs theories with TeV brane localized supersymmetry 
breaking. In this paper we explore theories where the two scales are 
tightly related through compactifications.

The boundary condition supersymmetry breaking in warped space 
has been considered before in Ref.~\cite{Gherghetta:2000kr}, but 
without addressing the issue of the consistency of the theory. 
Potential difficulties of the theory with supersymmetry broken 
by boundary conditions in a warped fifth dimension has been 
noted in Ref.~\cite{Gherghetta:2002nr}. Supersymmetry breaking 
boundary conditions were also considered in warped space 
in~\cite{Lalak:2002kx}.  In this paper we study the consistency 
of the theory in detail, and during that course we develop the concept 
of symmetry breaking defects in higher dimensional spacetime. In general, 
higher dimensional theories compactified on a spacetime with boundaries 
can possess symmetry breaking defects at the boundaries. When do such 
defects lead to consistent theories, and when do difficulties arise? 
In section~\ref{sec:defect} we study the local breaking of global 
internal symmetries in flat space, and introduce a distinction between 
two types of defect: type~I (type~II) defects which are (are not) 
consistent with a gauging of the global symmetry.  For example, 
we find that the boundary condition breaking of a $U(1)$ gauge symmetry, 
or of the electroweak gauge symmetry $SU(2)_L \times U(1)_Y \rightarrow 
U(1)_{EM}$, leads to type~II defects and thus is not consistent in 
a flat fifth dimension. In section~\ref{sec:symmetry} we show that the 
defects arising when supersymmetry is broken by a boundary condition 
in a warped fifth dimension are of type~II, preventing a consistent 
construction of the corresponding supergravity theory. Despite this 
difficulty, in section~\ref{sec:model} we construct an $SU(3)_C \times 
SU(2)_L \times U(1)_Y$ model in a warped 5D background with supersymmetry 
broken by boundary conditions. Such an effective theory may follow 
from some consistent fundamental theory. We explore electroweak symmetry 
breaking in this theory when there is a single Higgs hypermultiplet. 
We also construct an $SU(5)$ theory in warped 5D spacetime where 
supersymmetry is broken by boundary conditions in the fifth dimension, 
and show that consistent phenomenology is obtained in the theory.

\section{Symmetry Breaking Defects in Higher Dimensions}
\label{sec:defect}

In this section we carefully study the notion of symmetry breaking 
defects in higher dimensional effective field theories. These defects
arise on a boundary of the bulk when the Lagrangian at that boundary
is invariant under a smaller internal symmetry than that of the bulk
Lagrangian. We find that there are two types of defect: type~I defects
arise when the reduction in symmetry from bulk to boundary is entirely
forced by the boundary conditions imposed on the fields of the theory. 
On the other hand, for type~II defects not all of the symmetry 
reduction is required by the boundary conditions. For the first kind 
of defect, the internal symmetry can be gauged, and such defects were 
considered in Ref.~\cite{Hall:2001pg} in the context of higher 
dimensional grand unified theories. We discuss these defects in 
sub-section~\ref{subsec:type1-defect} using the example of $SU(5)$ 
symmetry in 5D.  In sub-section~\ref{subsec:type2-defect}, we introduce 
the second type of defect and find that the internal symmetry cannot
be consistently gauged. In section~\ref{sec:symmetry} we extend this 
analysis of defects to the case of supersymmetry, and find that 
it has important consequences for supersymmetry breaking in truncated 
AdS$_5$ space.

\subsection{Type~I symmetry breaking defect}
\label{subsec:type1-defect}

In this sub-section we discuss the first kind of symmetry breaking 
defect. This type of defect allows the whole symmetry structure 
to be gauged, and we call them type~I symmetry breaking defects. 
To illustrate the point, in this sub-section we consider 5D theories 
compactified on a flat $S^1/Z_2$ orbifold: a line segment 
parameterized by $y: [0, \pi R]$ with the metric of the spacetime 
given by $-ds^2 = g_{MN} dx^M dx^N = \eta_{\mu\nu} dx^\mu dx^\nu + dy^2$.

Let us first consider the theory in which the bulk Lagrangian possesses 
a global $SU(5)$ symmetry: for example, the bulk Lagrangian is
invariant under the transformation 
\begin{equation}
  \phi \rightarrow \exp(i T^A \xi^A)\phi, 
\label{eq:global}
\end{equation}
where the field $\phi$ is in the ${\bf 5}$ representation, $T^A$ are 
the generators of $SU(5)$ and $\xi^A$ are arbitrary constants. If the 
spacetime we consider were non-compact, this would be the end of the 
story.  However, since we are considering the theory on a compact 
space ($S^1/Z_2$ orbifold), we have to specify the boundary conditions 
on the fields to define the theory.  Suppose we require that all 
fields in a single irreducible representation of $SU(5)$ obey the same 
boundary conditions.  In this case the full theory can possess the 
global $SU(5)$ symmetry of Eq.~(\ref{eq:global}), and the resulting 
space does not have any symmetry breaking defect. What happens if 
we impose different boundary conditions on fields in the same 
irreducible representation of $SU(5)$?  This is the situation we 
want to consider in this sub-section.

The boundary conditions on $S^1/Z_2$ are completely specified if we 
specify the conditions which the fields must satisfy at $y=0$ and 
$y=\pi R$.  In general these conditions are written as
\begin{equation}
  \varphi(y) = \mbox{\bf{Z}}\,\varphi(-y), \qquad
  \varphi(y') = \mbox{\bf{Z}}'\,\varphi(-y'),
\label{eq:bc-fields}
\end{equation}
where $y' \equiv y-\pi R$; $\varphi$ is a column vector collecting 
all the fields in the theory, while $\mbox{\bf{Z}}$ and $\mbox{\bf{Z}}'$ 
are matrices acting on this vector.  The precise meaning of these 
conditions is the following.  Although our space is only for 
$0 \leq y \leq \pi R$, we can fictitiously extend it to the domain 
$y < 0$ or $y > \pi R$ using the above equations.  Then the dynamics 
of the fields (wavefunctions of the fields) are obtained by solving 
the equations of motion in the whole covering space, including the 
terms arising from brane-localized operators.  (The importance of 
thinking in this way becomes clearer in the next section because, 
unlike the flat space case, in AdS space we cannot construct the 
theory on $S^1/Z_2$ by simple identification procedures from the 
corresponding theory on the non-compactified AdS space.)

Now, we consider the matrices $\mbox{\bf{Z}}$ and/or $\mbox{\bf{Z}}'$ 
which do not give the same boundary conditions for all the fields 
in a single irreducible representation of $SU(5)$. For illustrative 
purposes, we choose these matrices to be $\mbox{\bf{Z}} = 
{\rm diag}(1,1,1,1,1)$ and $\mbox{\bf{Z}}' = {\rm diag}(-1,-1,-1,1,1)$ 
when acting on an $SU(5)$ fundamental index. For instance, the triplet 
and doublet components, $\phi_T$ and $\phi_D$, of the ${\bf 5}$ 
representation obey the boundary conditions $\phi_T(+,-)$ and 
$\phi_D(+,+)$, where the first and second signs represent the 
eigenvalues of $\mbox{\bf{Z}}$ and $\mbox{\bf{Z}}'$.

What are the consequences of imposing the above boundary conditions?
First of all, the whole theory obviously does not have a global $SU(5)$ 
symmetry, since we have imposed different boundary conditions on, say, 
$\phi_T$ and $\phi_D$ and they have different wavefunctions.  The
transformation of Eq.~(\ref{eq:global}) is inconsistent with the boundary
conditions at $y = \pi R$, again demonstrating the absence of the
global $SU(5)$ symmetry. However, physically we suspect that the physics
at any local neighborhood of the bulk must still reflect the original
global $SU(5)$ symmetry.  This is because the effect of the boundary 
conditions at $y=\pi R$, which is $SU(5)$ violating, is suppressed by 
locality in any point in the bulk.  On the other hand, at the $y=\pi R$ 
brane, the effect of $SU(5)$-violating boundary conditions is maximal, 
and we suspect that physics will not reflect the original global $SU(5)$ 
symmetry.  For example, the wavefunction values for $\phi_D$ can be 
non-zero at $y=\pi R$, while those for $\phi_T$ must always be zero. 
This implies that it does not make sense to impose the $SU(5)$ symmetry 
on the operators on the $y=\pi R$ brane. Hence we are led to ask: 
what is the most general form of the action, and is there a symmetry 
transformation which guarantees this form?

We find the most general form for the action to be
\begin{equation}
  S = \int\!\!d^4x \int\!\!dy 
    \Bigl[ {\cal L}_{\rm 5D}^{SU(5)} 
    + \delta(y) {\cal L}_{\rm 4D}^{SU(5)}
    + \delta(y-\pi R) {\cal L}_{\rm 4D}^{3-2-1} \Bigr].
\label{eq:Lag-SU5}
\end{equation}
Here, ${\cal L}_{\rm 5D}^{SU(5)}$ and ${\cal L}_{\rm 4D}^{SU(5)}$ 
respect the full $SU(5)$ symmetry, while ${\cal L}_{\rm 4D}^{3-2-1}$ 
respects only the $SU(3) \times SU(2) \times U(1)$ subgroup of
$SU(5)$. The different pieces of the Lagrangian are invariant under 
global transformations of different size:
\begin{eqnarray} 
  {\cal L}_{\rm 5D}^{SU(5)}, {\cal L}_{\rm 4D}^{SU(5)}: \;\;\; 
    \phi &\rightarrow& \exp(i T^A \xi^A)\phi,
\label{eq:restricted-1}\\
  {\cal L}_{\rm 4D}^{3-2-1}: \;\;\; 
    \phi &\rightarrow& \exp(i T^a \xi^a)\phi, 
\label{eq:restricted-2}
\end{eqnarray}
where $A$ runs over all $SU(5)$ generators while $a$ runs over the
subset of those forming the $SU(3) \times SU(2) \times U(1)$ subgroup,
and $\xi^A$ are constant and do not depend on the coordinates. 
This is an unusual situation --- while the theory does possess 
a global $SU(3) \times SU(2) \times U(1)$ symmetry, the other 
transformations of $SU(5)$ are not symmetries, since not all pieces 
of the Lagrangian are invariant under them. In general in higher 
dimensional theories, it is useful to consider an action where 
the bulk Lagrangian and the boundary Lagrangian possess different 
invariances. We will say that such theories possess restricted 
symmetries. Where the invariance at a boundary is less than 
in the bulk we will say that there is a symmetry breaking defect 
at the boundary. In our $SU(5)$ example, we therefore find that the 
boundary conditions have forced a reduction of the original global 
symmetry to a restricted global symmetry, with a symmetry breaking 
defect appearing at the $y=\pi R$ brane. The question is whether 
this new concept of a restricted global symmetry, such as 
Eqs.~(\ref{eq:restricted-1},~\ref{eq:restricted-2}), is really 
useful: does it lead to relations amongst counterterms, for example 
sufficient to yield Eq.~(\ref{eq:Lag-SU5}) as the most general action? 
Locality suggests that this is so: at short distances ({\it i.e.} with 
large momentum) in the bulk, the effect from the $y=\pi R$ boundary 
is exponentially suppressed due to Yukawa suppression (the 4D momentum 
appears as a mass in the direction of the fifth dimension).  The same 
argument applies to the Lagrangian at the $y=0$ brane.  Therefore, 
we expect that all divergences are absorbed into the counterterms 
preserving the form of Eq.~(\ref{eq:Lag-SU5}).

This expectation is confirmed because the theory defined by 
Eq.~(\ref{eq:Lag-SU5}) possesses a conserved $SU(5)$ current in the 
bulk, and a conserved $SU(3) \times SU(2) \times U(1)$ current at 
$y=\pi R$, at the quantum level. The notion of a restricted global 
symmetry, which takes a different form at different locations, makes 
sense because current conservation occurs locally. We can demonstrate 
that these currents are conserved, for instance, by the Noether 
procedure in the path integral formalism.  We consider varying the 
fields with position dependent $\xi$'s. The position dependence of 
$\xi$'s must be consistent with the boundary conditions of the fields 
and with the form of the restricted global symmetry. Specifically, 
we have to restrict the $y$ dependence of $\xi$'s as $\xi^a(+,+)$ and 
$\xi^{\hat{a}}(+,-)$ where $a$ and $\hat{a}$ run for $SU(3) \times 
SU(2) \times U(1)$ and $SU(5)/(SU(3) \times SU(2) \times U(1))$, 
respectively.  When expanded in the complete set in the fifth dimension, 
they are written as
\begin{eqnarray}
  \xi^a(x^\mu,y)
    &=& \sum_{n=0} \xi_n^a(x^\mu) \cos\left(\frac{n y}{R}\right),
\label{eq:su5-gtp-1}
\\
  \xi^{\hat{a}}(x^\mu,y)
    &=& \sum_{n=0} \xi_n^{\hat{a}}(x^\mu) 
    \cos\left(\frac{(n+1/2)y}{R}\right).
\label{eq:su5-gtp-2}
\end{eqnarray}
Note that with $\xi^{\hat{a}}$ having boundary conditions $(+,-)$ 
we automatically have $\xi^{\hat{a}}(x^\mu,y=\pi R)=0$, ensuring 
that we restrict transformations to be in $SU(3) \times SU(2) \times 
U(1)$ at $y=\pi R$.  The rest of the procedure is the usual one.
Although the action, Eq.~(\ref{eq:Lag-SU5}), is not invariant under 
the transformation by Eqs.~(\ref{eq:su5-gtp-1},~\ref{eq:su5-gtp-2}), 
the variation is proportional to the derivatives of $\xi$'s since 
Eq.~(\ref{eq:Lag-SU5}) is invariant under transformations with 
constant $\xi$'s.  This leads to a conservation law, which tells 
us that there is a conserved current for $SU(5)$ in the bulk and on 
$y=0$, but only the $SU(3) \times SU(2) \times U(1)$ part of it is 
conserved at $y=\pi R$.

We are now in a position to consider gauging the restricted global 
symmetry of the system.  It is the gauging which distinguishes
between the two types of defects discussed in this and the next 
sub-sections.  The gauging of the restricted global symmetry is 
accomplished by requiring the theory to be {\it invariant} under 
the transformations of Eqs.~(\ref{eq:su5-gtp-1},~\ref{eq:su5-gtp-2}) 
with arbitrary functions of $\xi_n^{a}(x^\mu)$ and 
$\xi_n^{\hat{a}}(x^\mu)$.\footnote{
The gauging is possible only when the theory is anomaly free. If the 
low energy 4D theory does not have anomalies, we can in general make 
the full higher dimensional theory to be anomaly free by introducing an 
appropriate Chern-Simons term in the bulk~\cite{Arkani-Hamed:2001is}.}
Since the kinetic terms of the original Lagrangian with restricted
global symmetry are not invariant under these transformations, 
we have to introduce the connection fields $A_M^A(x^\mu, y)$, 
which are in the adjoint representation of $SU(5)$.  The boundary 
conditions for these fields are determined to be $A_\mu^a(+,+)$ 
and $A_\mu^{\hat{a}}(+,-)$ ($A_5^a(-,-)$ and $A_5^{\hat{a}}(-,+)$) 
from the transformation properties of these fields, $A_M^A \rightarrow 
A_M^A + \partial_M \xi^A + \cdots$.  The expansion then goes 
as in Eqs.~(\ref{eq:su5-gtp-1},~\ref{eq:su5-gtp-2}) with $\xi^a$ 
and $\xi^{\hat{a}}$ replaced by $A_\mu^a$ and $A_\mu^{\hat{a}}$ (for 
$A_5$'s, replace the cosine by sine and start the sum for $A_{5,n}^a$ 
from $n=1$). Therefore, we find that there is a one-to-one correspondence 
between the modes $A_{\mu,n}^A$ and $\xi_n^A$.  This is crucial for 
the consistency of the gauge theory: each gauge field requires 
a corresponding gauge symmetry.  This one-to-one correspondence 
characterizes what we call type~I symmetry breaking defects. In our 
$SU(5)$ example, gauging produces a restricted gauge symmetry (the 
transformations of Eqs.~(\ref{eq:restricted-1},~\ref{eq:restricted-2}) 
with all $\xi$ now local) yielding a consistent effective higher 
dimensional field theory below the cutoff, as discussed in detail 
in Ref.~\cite{Hall:2001tn}. Restricted gauge symmetries play 
an important role for constructing realistic higher dimensional 
grand unified theories~\cite{Hall:2001pg}, which have automatic 
doublet-triplet splitting~\cite{Kawamura:2000ev}, proton decay 
suppression~\cite{Hall:2001pg}, and an interesting new prediction 
for gauge coupling unification~\cite{Hall:2001xb}.
In the next sub-section, we consider a different kind of defect, 
which violates the above one-to-one correspondence, and consequently 
does not allow the consistent gauging of the symmetry.

\subsection{Type~II symmetry breaking defect}
\label{subsec:type2-defect}

As in the previous sub-section, we consider a theory on the flat 
$S^1/Z_2$ orbifold.  We consider a restricted symmetry where the bulk and the 
$y=0$ brane possess a global $U(1)$ invariance but the $y=\pi R$ 
brane does not.  The action of this system takes the form:
\begin{equation}
  S = \int\!\!d^4x \int\!\!dy 
    \Bigl[ {\cal L}_{\rm 5D}^{U(1)} 
    + \delta(y) {\cal L}_{\rm 4D}^{U(1)}
    + \delta(y-\pi R) {\cal L}_{\rm 4D}^{\times} \Bigr].
\label{eq:Lag-U1}
\end{equation}
Here ${\cal L}_{\rm 5D}^{U(1)}$ and ${\cal L}_{\rm 4D}^{U(1)}$ 
are invariant under the field rotation $\phi \rightarrow 
\exp(i Q_\phi \xi)\phi$, but ${\cal L}_{\rm 4D}^{\times}$ is not, where 
$\phi$ is a field carrying the $U(1)$ charge of $Q_\phi$ and $\xi$ is 
an arbitrary constant.  The boundary conditions for $\phi$ are taken 
to be either $(+,+)$, $(+,-)$, $(-,+)$ or $(-,-)$ [the other 
possibilities are mentioned in footnote~2].

Does the above action make sense?  To answer this question, we have 
to study radiative corrections.  As in the previous example, we 
find that all divergences are absorbed in the counterterms preserving 
the form of the Lagrangian.  Here we prove this using the Noether 
procedure in the path integral formalism.  We consider the $U(1)$ 
transformation parameter $\xi$ to be a function of the spacetime.
The boundary conditions for $\xi$ are determined to be $\xi(+,+)$ so 
that this $U(1)$ transformation preserves the boundary conditions for 
$\phi$'s.  A mode expansion gives 
\begin{equation}
  \xi(x^\mu,y)
    = \sum_{n=0} \xi_n(x^\mu) \cos\left(\frac{n y}{R}\right).
\label{eq:u1-gtp}
\end{equation}
However, we now have an extra constraint.  Because the above expansion 
does not ensure the vanishing of $\xi(x^\mu,y)$ at $y=\pi R$, where 
$U(1)$ symmetry is supposed to be absent, we have to impose a further
condition on the $\xi_n^a(x^\mu)$'s:
\begin{equation}
  \sum_{n=0} \xi_n(x^\mu) \cos(\pi n) = 0.
\label{eq:u1-gtp-const}
\end{equation}
We now vary the action with arbitrary $\xi_n^a(x^\mu)$'s under the 
constraint Eq.~(\ref{eq:u1-gtp-const}).  Then we find that the variation 
is proportional to the derivative of $\xi(x^\mu,y)$, giving a current 
associated with $U(1)$ which is conserved everywhere except $y=\pi R$.
Therefore, we find the system with a restricted global $U(1)$ 
symmetry with a symmetry breaking defect at $y=\pi R$ is meaningful, 
in the sense that its structure is preserved by radiative corrections.
This situation is quite similar to the restricted global $SU(5)$
symmetry with the $SU(3) \times SU(2) \times U(1)$ defect.

Now, we consider gauging this restricted $U(1)$ global symmetry, 
{\it i.e.} we require the theory to be {\it invariant} under 
position dependent $\xi$.  To make the kinetic term of the original 
Lagrangian invariant, we must introduce the connection fields, 
$A_M(x^\mu,y)$.  The boundary conditions for these fields are 
determined to be $A_\mu(+,+)$ and $A_5(-,-)$ from their transformation 
properties $A_M \rightarrow A_M + \partial_M \xi$.  Therefore, the 
mode expansions for these fields are given by Eq.~(\ref{eq:u1-gtp}) 
with $\xi$ replaced by $A_\mu$ (for $A_5$, replace the cosine by sine 
and start the sum from $n=1$).  Unlike the gauge parameter $\xi$, 
however, these gauge fields $A_M$ are dynamical fields, so that we 
cannot simply impose the constraint like Eq.~(\ref{eq:u1-gtp-const}). 
In particular, all $A_{\mu,n}(x^\mu)$ are independent fields. 
This means that the number of gauge transformation parameters, 
$\xi_n(x^\mu)$, is smaller than the number of gauge fields, 
$A_{\mu,n}(x^\mu)$, due to the constraint imposed on the $\xi_n(x^\mu)$'s, 
Eq.~(\ref{eq:u1-gtp-const}).  This leads to an inconsistency of 
the theory, because, from the 4D viewpoint, there is one gauge field 
which is not accompanied by a corresponding gauge symmetry. 
As is well known, such a gauge field gives a ghost which can be 
produced as an external particle, leading to negative probabilities 
for certain processes.

Therefore, we find that the defect in this $U(1)$ theory has a different 
character from the one discussed in the previous sub-section, and 
we call it a defect of type~II. When the restricted symmetry is global 
the two types of defect have similar properties, but when the restricted 
symmetry is gauged quite different features are revealed: one 
allows consistent gauging but the other does not. The criterion for 
distinguishing the two is whether the number of gauge transformation 
parameters is the same as or smaller than the number of the gauge 
fields (counting modes in the 4D picture).  Type~I defects arise 
naturally when the restricted symmetry is taken to be the largest 
possible consistent with the boundary conditions imposed on the fields. 
In the $U(1)$ example, the boundary conditions are consistent with 
a $(+,+)$ parity assignment to $\xi$, so that both branes would 
naturally be expected to have Lagrangians which respect the $U(1)$ 
symmetry.\footnote{
If the theory possesses a charge-conjugation symmetry, say has 
two fields $\varphi$ and $\bar{\varphi}$ with the opposite charges 
$Q_{\varphi} = -Q_{\bar{\varphi}}$, one can choose the boundary 
conditions at $y=\pi R$ so that all the fields are identified with 
the corresponding charge-conjugated ones, $\varphi(x^\mu,y) = 
\bar{\varphi}(x^\mu,y)|_{y=\pi R}$, which forces $\xi(x^\mu,y)$ to 
be vanishing at $y=\pi R$ (if $\varphi$ is a scalar field, we could 
choose $\varphi(x^\mu,y) = \varphi^*(x^\mu,y)|_{y=\pi R}$ and do 
not necessarily need $\bar{\varphi}$). In this case the resulting 
$U(1)$-breaking defect at $y=\pi R$ is type~I, because it is the 
largest possible symmetry consistent with the boundary conditions 
imposed on the fields. The transformation properties for the gauge 
field and transformation parameter are then given by $A_\mu(+,-)$, 
$A_5(-,+)$ and $\xi(+,-)$, assuming the usual Neumann or Dirichlet 
boundary conditions for $\varphi$ and $\bar{\varphi}$ at $y=0$.}
Type~II defects arise when the invariance on the boundary 
is taken to be less than the maximal consistent with the boundary 
conditions. Clearly there are much more general possibilities than 
we have discussed, even on $S^1/Z_2$. The restricted global symmetry 
may correspond to invariances of the three pieces of the Lagrangian 
under transformations of different sizes. Type~I defects arise if 
these transformations on a boundary Lagrangian are the largest
consistent with the boundary conditions that have been imposed on 
the fields. If the boundary Lagrangian is invariant under a smaller 
set of transformations, then the defect is type~II. 

A case of potential phenomenological interest has the bulk Lagrangian 
invariant under the electroweak group $SU(2)_L \times U(1)_Y$, with 
a boundary Lagrangian invariant under the smaller electromagnetic 
symmetry group $U(1)_{EM}$. Apparently this provides an alternative 
electroweak symmetry breaking mechanism --- no Higgs bosons or new 
strong dynamics are needed since the defect explicitly breaks weak 
interactions. However, this defect is of type~II: there is no way of 
imposing boundary conditions on the fields such that the reduction of 
symmetry on the boundary is required for consistency with the boundary 
conditions. Incidentally, in the next section when we consider boundary 
condition breaking of supersymmetry in warped space, we will similarly 
discover that the defects are of type~II.

Although we find that type~II defects do not allow gauging of the 
symmetry in a straightforward way, we can obtain a low energy
effective field theory which mimics the gauging of type~II defects. 
For instance, to mimic the above theory, we can first consider 
a 5D $U(1)$ gauge theory compactified on the flat $S^1/Z_2$ orbifold 
without any defect.  Then, if we break this $U(1)$ by the vacuum 
expectation value for the Higgs field $h$ localized on the $y=\pi R$ 
brane, we find that the wavefunctions for the gauge field are given 
by $\sim \cos((n+1/2)y/R)$ in the limit $\langle h \rangle \rightarrow 
\infty$~\cite{Nomura:2001mf}.  The operators on the $y=\pi R$ 
brane can now pick up the effect of this large expectation value, 
so that they effectively do not respect the $U(1)$ symmetry. 
Thus the action of the resulting effective field theory is given 
by Eq.~(\ref{eq:Lag-U1}) with the $U(1)$ symmetry gauged.
Although this theory with large brane Higgs expectation value does 
not completely reproduce the properties of the theory where the 
restricted global symmetry with type~II defects were gauged, it shares 
many properties with such a (non-existent) theory.  Therefore, it 
may not be so meaningless to consider theories with type~II defects 
and consider the gauging of its symmetry, in the sense that we might 
find some underlying theory reproducing some of the features 
possessed by such a theory.  This is the attitude we will take 
in section~\ref{sec:model} when we consider theories with boundary 
condition supersymmetry breaking in warped space.

\section{Supersymmetry Breaking in Warped Space}
\label{sec:symmetry}

In this section we study the supersymmetry structure of theories on 
truncated AdS$_5$ space, {\it i.e.} AdS$_5$ with the fifth 
dimension compactified on the $S^1/Z_2$ orbifold.  The metric for 
5D AdS space with 4D Poincare invariance is given by
\begin{equation}
  -ds^2 = g_{MN} dx^M dx^N = 
    e^{-2\sigma(y)}\eta_{\mu\nu} dx^\mu dx^\nu + dy^2,
\label{eq:AdSmetric}
\end{equation}
where $\eta_{\mu\nu}=\mbox{diag}(-1,1,1,1)$ and $\sigma(y)=ky$ with 
$0 \leq y \leq \pi R$.

Although the physical space of $S^1/Z_2$ is only $0 \leq y \leq \pi R$, 
we can extend it to all values for $y$ with the understanding that 
different points are identified as $y \sim -y$ and $y' \sim -y'$, where 
$y' = y-\pi R$.  This is a useful procedure because we can then figure 
out the physics at the boundaries, $y=0,\pi R$, just by considering the 
equations of motion {\it etc.} across these points. The extension of 
the metric to the (fictitious) space $y<0$ and $y>\pi R$ is given by 
\begin{equation}
  \left\{ \begin{array}{l}
    \sigma(y)=k|y| ~~\mbox{in}~ -\pi R \leq y \leq\pi R, \\
    \sigma(y+2\pi R)=\sigma(y),
  \end{array} \right.
\label{eq:sigma-orb}
\end{equation}
since $g_{\mu\nu}$ must be even under $y \rightarrow -y$ and $y' 
\rightarrow -y'$.

In sub-section~\ref{subsec:bulk} we define global supersymmetry 
in AdS space and write down the bulk Lagrangian.  The effects 
of the boundaries are considered in sub-section~\ref{subsec:boundary}.
We show that, if we impose boundary conditions on the fields that 
preserve $N=1$ supersymmetry in 4D, the two boundaries at $y=0$ and 
$\pi R$ are supersymmetry breaking defects of type~I in the 
classification of the previous section.  On the other hand, if 
we impose boundary conditions which break all the supersymmetries 
in 4D, we find that the resulting defect is type~II.  This leads 
to an important result that when we gauge supersymmetry, which is 
required to incorporate gravity into the theory, then the theory 
with supersymmetry breaking boundary conditions becomes inconsistent. 
Therefore, if we want to consider warped theories with supersymmetry 
breaking boundary conditions, such theories must be viewed, at best, 
as phenomenological approximations to some consistent theories that 
mimic the desired properties of the theories with boundary condition 
supersymmetry breaking.

\subsection{Supersymmetry in the bulk of AdS$_5$}
\label{subsec:bulk}

In this sub-section we study supersymmetry in AdS space and write 
down the off-shell Lagrangian in the bulk of $S^1/Z_2$.  Recall that 
a commutator of two supersymmetry transformations $\delta_\xi$ and 
$\delta_\eta$, parameterized by two Dirac spinors $\xi$ and $\eta$ 
respectively,\footnote{
We use the following convention for $\gamma$-matrices:
\begin{equation*}
  \{\gamma^M,\gamma^N\}=2g^{MN},~~~
  \gamma^\mu = -i e^{\sigma(y)} 
    \left( \begin{array}{cc} 
      0 & \sigma^\mu \\ \bar{\sigma}^\mu & 0
    \end{array} \right)
    \equiv e^{\sigma(y)} \hat{\gamma}^\mu,~~~
  \gamma^5 = \gamma_5 = 
    -i\hat{\gamma}^0\hat{\gamma}^1\hat{\gamma}^2\hat{\gamma}^3
    = \left( \begin{array}{cc} 
      1 & 0 \\ 0 & -1
    \end{array} \right),
\end{equation*}
where $\sigma^\mu=(1,\vec{\sigma})$ and 
$\bar{\sigma}^\mu=(1,-\vec{\sigma})$.  The Dirac conjugate 
is defined as $\bar{\Psi} \equiv \Psi^\dagger i\hat{\gamma}^0$.}
acts on the coordinates $x^M$ as
\begin{eqnarray}
  x^M &\longrightarrow& x^M + \epsilon^M, 
\nonumber\\
  &\mbox{where}& [\delta_\eta,\delta_\xi] 
    = 2(\bar{\eta} \gamma^M \xi -\bar{\xi} \gamma^M \eta)\partial_M 
    \equiv \epsilon^{M}\partial_M .
\label{eq:susycomm}
\end{eqnarray}
Under this coordinate transformation, the metric $g_{MN}$ changes as
\begin{equation}
  g_{MN} \longrightarrow g_{MN} + \epsilon^L \partial_L g_{MN}
    + g_{LN} \partial_M \epsilon^L + g_{ML} \partial_N \epsilon^L.
\end{equation}

Now, a global supersymmetry transformation is defined as the 
supersymmetry transformation which leads to $\epsilon^M$ that leaves 
$g_{MN}$ unchanged. Namely, we require $\epsilon^M$ to satisfy
\begin{equation}
  \epsilon^L \partial_L g_{MN}
    + g_{LN} \partial_M \epsilon^L + g_{ML} \partial_N \epsilon^L = 0,
\end{equation}
or more explicitly
\begin{eqnarray}
  && \partial_5 \epsilon^5 = 0,
\label{eq:killingvector1} \\
  && g_{\mu\nu}\partial_5\epsilon^\nu 
     + \partial_\mu\epsilon^5 = 0,
\label{eq:killingvector2} \\
  && -2\sigma' g_{\mu\nu} \epsilon^5 
     + g_{\rho\nu}\partial_\mu\epsilon^\rho
     +g_{\mu\rho}\partial_\nu\epsilon^\rho = 0,
\label{eq:killingvector3}
\end{eqnarray}
where $\sigma' \equiv \partial\sigma/\partial y$.  The vector 
$\epsilon^M$ is called a Killing vector, and the above equations are 
called Killing vector equations.

By replacing $\epsilon^M$ in 
Eqs.~(\ref{eq:killingvector1}~--~\ref{eq:killingvector3}) by 
Eq.~(\ref{eq:susycomm}), we find that the Killing vector equations 
are satisfied if $\xi$ (and $\eta$) satisfies certain conditions. 
Such a spinor is called a Killing spinor.  We write these conditions, 
called Killing spinor equations, using the symplectic Majorana 
spinor notation: we express the 5D supersymmetry transformation 
parameter $\xi$ by two Dirac spinors $\xi^1$ and $\xi^2$ obeying 
a single relation.\footnote{
Here, $\xi^1$ and $\xi^2$ {\it together} correspond to a {\it single} 
Dirac spinor $\xi$. They are related as
\begin{equation*}
  \xi^1 \equiv \xi, ~~~ \xi^2 \equiv -C\xi^*,
  \quad \mbox{so that} ~~ \xi^i = \epsilon^{ij}C\xi^*_j 
  ~~ \mbox{and} ~~ \xi^*_i = (\xi^i)^*,
\end{equation*}
where $C \equiv -\hat{\gamma^2}\gamma^5$ is the 5D charge conjugation 
matrix and has properties, $C^2=-1$ and $C \gamma^M C^{-1} 
= -\gamma^{M*}$. Thus both $\xi^1$ and $\xi^2$ properly transform 
as 5D Dirac spinors, and simultaneously they form a doublet under the 
$SU(2)_R$ automorphism group of the 5D supersymmetry. In terms of more 
familiar two component notation, they are:
\begin{equation*}
  \xi^1 = \left(
    \begin{array}{c} 
      \xi_{L\alpha} \\ \bar{\xi}_R^{\dot{\alpha}}
    \end{array} \right), 
~~~ 
  \xi^2 = \left(
    \begin{array}{c}
      -\xi_{R\alpha} \\ \bar{\xi}_L^{\dot{\alpha}}
    \end{array} \right).
\end{equation*}
There is one convenient identity for these spinors: 
$\bar{\xi}_i\gamma^M\cdots\gamma^K\eta^j = 
\bar{\eta}^j\gamma^K\cdots\gamma^M\xi_i$, for any 
$\xi^i$ and $\eta^i$.}
In this notation, Eq.~(\ref{eq:susycomm}) simply becomes 
$\epsilon^M=\bar{\eta}_i\gamma^M\xi^i$. First, we find 
that the most general form for the constraint that solves 
Eq.~(\ref{eq:killingvector1}) and is consistent with the 
4D Lorentz invariance is given by
\begin{equation}
  \partial_5 \xi^i 
    = -\frac{\sigma'}{2} H^i_j \gamma_5\xi^j
      -\frac{i\sigma'}{2} K^i_j \xi^j,
\label{eq:killingspinoransatz1}
\end{equation}
where $\xi^i$ represents a general Killing spinor, and $H^i_j$ and 
$K^i_j$ are $2\times2$ arbitrary Hermitian matrices which can even 
depend on positions in spacetime. At this stage, the only constraints 
for these matrices come from differentiating the identity 
$\xi^i=\epsilon^{ij}C(\xi^j)^*$ by $y$, which leads to
\begin{equation}
  \mbox{Tr}[H] = \mbox{Tr}[K] = 0.
\end{equation}
We next consider Eq.~(\ref{eq:killingvector2}) and find that, in order 
to solve this, we need an equation for $\partial_\mu\xi^i$ as well as 
Eq.~(\ref{eq:killingspinoransatz1}).  The most general form of this is
given by
\begin{equation}
  \partial_\mu\xi^i 
    = -\frac{\sigma'}{2} H^i_j\gamma_\mu\xi^j
      -\frac{\sigma'}{2} \gamma_5\gamma_\mu\xi^i
      -\frac{i\sigma'}{2} L^i_j\gamma_\mu\xi^j,
\label{eq:killingspinoransatz2}
\end{equation}
where $L^i_j$ is a new arbitrary $2\times2$ Hermitian matrix. Finally, 
we consider the last equation, Eq.~(\ref{eq:killingvector3}). We find 
that this equation is satisfied if and only if
\begin{equation}
  L^i_j = 0.
\end{equation}
Therefore, we find that the Killing spinor must satisfy the equations
\begin{eqnarray}
  \partial_5 \xi^i 
  &=& -\frac{\sigma'}{2} H^i_j \gamma_5\xi^j
      -\frac{i\sigma'}{2} K^i_j \xi^j,
\label{eq:naivekillingspinorequ1} \\
  \partial_\mu \xi^i 
  &=& -\frac{\sigma'}{2} H^i_j\gamma_\mu\xi^j
      -\frac{\sigma'}{2} \gamma_5 \gamma_\mu \xi^i,
\label{eq:naivekillingspinorequ2}
\end{eqnarray}
where $H^i_j$ and $K^i_j$ are arbitrary $2\times2$ traceless Hermitian 
matrices.

Let us now examine whether 
Eqs.~(\ref{eq:naivekillingspinorequ1},~\ref{eq:naivekillingspinorequ2}) 
have a non-trivial solution or not.  If there exists a non-trivial 
and reasonable $\xi$, it must satisfy
\begin{equation}
  [\partial_M, \partial_N] \xi^i = 0.
\end{equation}
Evaluating the above commutator for $M=\mu$ and $N=\nu$ gives the 
following constraints on the matrix $H$:
\begin{eqnarray}
  H^2 &=& {\bf 1},
\label{eq:const-1} \\
  \partial_\mu H &=& {\bf 0},
\label{eq:const-2}
\end{eqnarray}
where ${\bf 1}$ and ${\bf 0}$ are the unit and zero $2 \times 2$ 
matrices.  On the other hand, $[\partial_\mu,\partial_5]\xi^i=0$ gives
the following constraints on $H$ and $K$:
\begin{eqnarray}
  \partial_\mu K &=& {\bf 0},
\label{eq:const-3} \\
  -i\partial_5 H &=& \left[ -\frac{\sigma'}{2}K, H \right],
\label{eq:const-4}
\end{eqnarray}
and the conditions for $\xi^i$:
\begin{equation}
  H^i_j(y) \xi^j = \gamma_5 \xi^i \qquad 
  \mbox{at } y=0 \mbox{ and } \pi R.
\label{eq:basicboundarycondition}
\end{equation}
Because the form of Eq.~(\ref{eq:const-4}) is identical to the 
Heisenberg equation of motion for the operator $H$ with ``time" $y$ 
and ``Hamiltonian" $-(\sigma'/2)K$, we can write the general 
solution as:
\begin{equation}
   H(y) = U(y) H(0) U^\dagger(y),
\end{equation}
where, having Eqs.~(\ref{eq:const-1},~\ref{eq:const-3}) in mind, $H(0)$ and 
$U(y)$ are given by
\begin{eqnarray}
  H(0) &=& n^a \sigma_a,
\\
  U(y) &=& \hat{\mbox{{\bf Y}}} 
    \exp\left[ -\frac{i}{2} \int^y_0 \sigma' K(y') dy' \right],
\end{eqnarray}
where $n^a$ ($a=1,2,3$) is a constant real vector with unit 
length $n^an^a=1$, $\sigma_a$ are the Pauli spin matrices, and 
$\hat{\mbox{{\bf Y}}}$ is the ``time"-ordering operator. We can 
check that this solution solves all the constraints on $H$ and 
$K$, Eqs.~(\ref{eq:const-1}~--~\ref{eq:const-4}).  Note also that
$U(y)$ belongs to $SU(2)$ because $K$ is Hermitian and traceless. 

The above Killing spinor equation contains important information 
about the symmetry structure of the theory.  We consider the 
$SU(2)_R$ automorphism group of the 5D supersymmetry, under which 
$\xi^1$ and $\xi^2$ form a doublet.  In flat space ($\sigma' = 0$), 
this $SU(2)_R$ is a symmetry of the algebra and thus respected 
by the whole theory.  In AdS, however, we find that it 
is broken by the presence of the matrices $H$ and $K$ in 
Eqs.~(\ref{eq:naivekillingspinorequ1},~\ref{eq:naivekillingspinorequ2}). 
Now, we consider redefining the fields by a twist inside $SU(2)_R$. 
This results in the redefinition of $\xi^i$ according to
\begin{equation}
   \xi^i(y) \longrightarrow \tilde{U}(y)^i_j \xi^j(y),
\label{eq:redef}
\end{equation}
where $\tilde{U}(y)$ is a $y$-dependent matrix taking arbitrary 
values in $SU(2)$. Note that, since we are just redefining 
the name of the fields, this does not change any physics. 
Then, substituting Eq.~(\ref{eq:redef}) into 
Eqs.~(\ref{eq:naivekillingspinorequ1},~\ref{eq:naivekillingspinorequ2}) 
and choosing $\tilde{U}(y) = U(y)$, we find that $H(y)$ is replaced 
by $H(0)$ and the $K(y)$ term is canceled.  We can further make 
a $y$-independent $SU(2)_R$ rotation and choose $(n^1,n^2,n^3) = (0,0,1)$ 
for $H(0)$. Therefore, we finally obtain the following simple form for 
the Killing spinor equations in AdS$_5$:
\begin{eqnarray}
  \partial_5 \xi^i 
  &=& -\frac{\sigma'}{2} (\sigma_3)^i_j \gamma_5 \xi^j,
\label{eq:killingspeq-1} \\
  \partial_\mu \xi^i 
  &=& -\frac{\sigma'}{2} (\sigma_3)^i_j \gamma_\mu \xi^j 
     - \frac{\sigma'}{2} \gamma_5 \gamma_\mu \xi^i.
\label{eq:killingspeq-2}
\end{eqnarray}
These equations show that a $U(1)_R$ subgroup of $SU(2)_R$ remains 
unbroken in the AdS background.  The constraint on $\xi^i$, 
Eq.~(\ref{eq:basicboundarycondition}), now becomes
\begin{equation}
  (\sigma_3)^i_j \xi^j = \gamma_5 \xi^i \qquad 
  \mbox{at } y=0 \mbox{ and } \pi R.
\label{eq:xi-cond}
\end{equation}
These three equations define global supersymmetry in the truncated 
AdS$_5$ on $S^1/Z_2$. The form of the bulk Lagrangian is determined by
Eqs.~(\ref{eq:killingspeq-1},~\ref{eq:killingspeq-2}). 
The Killing spinor boundary constraint of Eq.~(\ref{eq:xi-cond}) 
is crucially important when we consider the effect of the boundaries 
in the next sub-section. In particular it requires that 
$\xi_{R\alpha} = 0$ at both boundaries.

Finally, we write down the off-shell bulk Lagrangians (in 
the basis where the Killing spinor equations take the form of
Eqs.~(\ref{eq:killingspeq-1},~\ref{eq:killingspeq-2})).
Effects of boundaries, including Eq.~(\ref{eq:xi-cond}), will be 
considered in the next sub-section.  We begin with a hypermultiplet, 
which consists of two complex scalars, $\phi^1$ and $\phi^2$, and 
a Dirac spinor, $\Psi$, and two complex auxiliary fields, $F^1$ 
and $F^2$. The kinetic part of the action is given by
\begin{equation}
  {\cal S}_{\rm hyp.kin.} 
    \equiv \int\!\!d^4x \int\!\!dy \, \sqrt{-g} \, {\cal L}_{\rm hyp.kin.},
\label{eq:def-s}
\end{equation}
\begin{equation}
  {\cal L}_{\rm hyp.kin.}
    = -g^{MN} \partial_M\phi^*_i \partial_N\phi^i
      -\frac12 \bar{\Psi} \gamma^M \partial_M\Psi
      +\frac12 \partial_M\bar{\Psi} \gamma^M \Psi
      +F^*_iF^i + \frac{15}{4}k^2 \phi^*_i\phi^i,
\label{eq:hyperaction}
\end{equation}
where $i=1,2$ and both $\phi^i$ and $F^i$ are doublets under $SU(2)_R$; 
in particular $\phi_i=\epsilon_{ij}\phi^j$, $\phi^*_i=(\phi^i)^*$, 
and so on. This action is invariant under the following global 
supersymmetry transformation:
\begin{eqnarray}
  && \delta\phi^i 
    = \sqrt2 \epsilon^{ij}\bar{\xi}_j\Psi, 
\nonumber\\
  && \delta\Psi 
    = \sqrt2 \left( \gamma^M \xi^i \partial_M\phi_i 
        - \frac32 \sigma' \xi^i \phi_j (\sigma_3)^j_i 
        + \xi^i F_i \right),
\label{eq:hypertrans}\\
  && \delta F^i 
    = \sqrt2 \epsilon^{ij}\left( \bar{\xi}_j \gamma^M \partial_M\Psi 
        - 2\sigma' \bar{\xi}_j \gamma_5 \Psi \right),
\nonumber
\end{eqnarray}
where the global supersymmetry transformation parameter $\xi^i$ 
satisfies Eqs.~(\ref{eq:killingspeq-1},~\ref{eq:killingspeq-2}).
In addition to the above kinetic part, Eq.~(\ref{eq:hyperaction}), 
we can also add a mass to the hypermultiplet:
\begin{equation}
  {\cal L}_{\rm hyp.mass} 
  = - c \sigma' \bar{\Psi} \Psi 
    + c \sigma' (F^*_i\phi^i + \phi^*_iF^i)
    - ck^2 (\sigma_3)^i_j \phi^*_i \phi^j,
\label{eq:hypermass}
\end{equation}
where $c$ is a dimensionless real constant. This by itself 
is invariant (up to a total derivative) under the global 
supersymmetry transformation, Eqs.~(\ref{eq:hypertrans}) with 
Eqs.~(\ref{eq:killingspeq-1},~\ref{eq:killingspeq-2}).

The gauge supermultiplet consists of a vector field $A_M$, a Dirac 
gaugino $\Psi_\lambda$, a real scalar $\Sigma$, and three real 
auxiliary fields $X^a$ ($a=1,2,3$). The Lagrangian is given by
\begin{eqnarray}
  {\cal L}_{\rm gauge}
    &=& \frac{1}{g^2} \biggl[ -\frac{1}{4}g^{ML}g^{NK}F_{MN}F_{KL}
      -\frac12 g^{MN} \partial_M\Sigma \> \partial_N\Sigma
      -\frac12 \bar{\lambda}_i \gamma^M \partial_M\lambda^i
      +\frac12 X^a X^a 
\nonumber\\
  && {} +2k^2 \Sigma^2 
      -\frac14 \sigma' (\sigma_3)^i_j \bar{\lambda}_i \lambda^j \biggr],
\label{eq:gaugelag}
\end{eqnarray}
where we have chosen the gauge group to be $U(1)$ for simplicity.
We have also used the symplectic Majorana notation for the gaugino: 
$\Psi_\lambda$ is represented by the two Dirac spinors $\lambda^1$ 
and $\lambda^2$.  Note that the gaugino and the auxiliary fields 
form a doublet and a triplet, respectively, under $SU(2)_R$. 
This Lagrangian is supersymmetric under the following global 
supersymmetry transformation:
\begin{eqnarray}
  && \delta A_M = -\bar{\xi}_i\gamma_M\lambda^i,
\nonumber\\
  && \delta \Sigma = i \bar{\xi}_i\lambda^i,
\nonumber\\
  && \delta \lambda^i = -i\gamma^M\xi^i \partial_M\Sigma
    -\frac12 \gamma^M\gamma^N \xi^i F_{MN}
    +2i\sigma'\Sigma(\sigma_3)^i_j\xi^j
    -i(\sigma_a)^i_j \xi^j X^a,
\label{eq:gaugetrans}\\
  && \delta X^a = i\bar{\xi}_i\gamma^M\partial_M\lambda^j (\sigma^a)^i_j
    -2i\sigma' \bar{\xi}_i\gamma_5\lambda^j (\sigma^a)^i_j,
\nonumber
\end{eqnarray}
where $\xi$ satisfies the condition 
Eqs.~(\ref{eq:killingspeq-1},~\ref{eq:killingspeq-2}).
Generalization to a non-Abelian group is fairly straightforward 
(giving the appropriate gauge structure, adding certain 
gaugino-gaugino-scalar interactions, changing the derivatives 
to gauge covariant derivatives, and so on).  These bulk Lagrangians, 
Eqs.~(\ref{eq:hyperaction},~\ref{eq:hypermass},~\ref{eq:gaugelag}), 
reproduce the on-shell bulk Lagrangians given in 
Refs.~\cite{Shuster:1999zf}, after integrating out the auxiliary 
fields (assuming no boundaries).

\subsection{Effects of the boundaries}
\label{subsec:boundary}

In this sub-section we consider the effects of the boundaries.
We follow the discussion in section~\ref{sec:defect} and consider 
the symmetry structure of the theory.  A new ingredient compared 
with the previous case is the constraint coming from the Killing 
spinor equation, Eq.~(\ref{eq:xi-cond}).  This additional complication 
arises from the fact that supersymmetry is a spacetime symmetry. 
The other parts of the discussion, however, are quite analogous 
to the previous case.

We begin by considering the boundary conditions on the fields. 
As explained in the previous section, the boundary conditions are 
written as Eq.~(\ref{eq:bc-fields}), where $\varphi$ is a column 
vector collecting all the fields in the theory, including the metric 
$g_{MN}$.  Since the matrices $\mbox{\bf{Z}}$ and $\mbox{\bf{Z}}'$ must 
be representations of the two reflections ${\cal Z}: y \rightarrow -y$ 
and ${\cal Z}': y' \rightarrow -y'$, respectively, they must obey 
the relations:
\begin{equation}
  \mbox{\bf{Z}}^{2} = {\bf 1}, \qquad
  \mbox{\bf{Z}}^{\prime 2} = {\bf 1}.
\label{eq:zz-z'z'}
\end{equation}
Thus we find that the general boundary conditions are given as 
follows. Under the reflection ${\cal Z}$, the fields obey 
\begin{eqnarray}
  && \phi^i(y) = P_\Phi U^i_j (\sigma_3)^j_k \phi^k(-y),
\nonumber\\
  && \Psi(y) = P_\Phi \gamma_5 \Psi(-y),
\label{eq:hyperparity}\\
  && F^{i}(y) = P_\Phi U^{i}_j (\sigma_3)_k^j F^k(-y),
\nonumber
\end{eqnarray}
and
\begin{eqnarray}
  && A_\mu(y) = A_\mu(-y), \qquad A_5(y) = -A_5(-y),
\nonumber\\
  && \lambda^i(y) = U^i_j (\sigma_3)^j_k \gamma_5 \lambda^k(-y),
\nonumber\\
  && \Sigma(y) = -\Sigma(-y),
\label{eq:gaugeparity}\\
  && X^a(y) = \frac12{\rm tr}
    [\sigma^aU\sigma^3\sigma^b\sigma^3U^\dagger]X^b(-y),
\nonumber
\end{eqnarray}
where $U = \exp[2\pi i(\alpha_1\sigma_1+\alpha_2\sigma_2)]$ with $0 \leq 
\alpha_{1,2} < 1$, and each hypermultiplet can have its own 
parity $P_\Phi = \pm 1$.  The boundary conditions under ${\cal Z}'$ 
is also given similarly, by the replacement $y \rightarrow y'$, 
$U \rightarrow U'$ ($\alpha_{1,2} \rightarrow \alpha'_{1,2}$) 
and $P_\Phi \rightarrow P'_\Phi$ in 
Eqs.~(\ref{eq:hyperparity},~\ref{eq:gaugeparity}), 
where $0 \leq \alpha'_{1,2} < 1$ and $P'_\Phi = \pm 1$.\footnote{
Here we have assumed the boundary conditions do not break the gauge
symmetry, although including such breaking is straightforward. The 
procedure is exactly identical to that in flat space.}

Now, we study the supersymmetry structure of the theory: 
a conservation law for the supercurrent. Following the discussion 
in section~\ref{sec:defect}, we consider the Noether procedure 
in the path integral formalism. What position dependence should 
we allow for the supersymmetry transformation parameter, and 
how many supersymmetries are preserved in each point in the extra 
dimension?  First, we can easily see that there are 4D $N=2$ 
supersymmetries in the bulk, because in any local neighborhood 
of the bulk we can solve the Killing spinor equation, 
Eqs.~(\ref{eq:killingspeq-1},~\ref{eq:killingspeq-2}), as 
$\xi^1(x^\mu,y) = \exp(-\sigma\gamma_5/2) (1 - \sigma'\exp(\sigma)
\gamma_\mu x^\mu (1-\gamma_5)/2) \xi_0$, which is parameterized by 
an arbitrary constant Dirac spinor $\xi_0$. A non-trivial question 
is the number of supersymmetries on the boundaries.  At the 
boundaries $y=0$ and $\pi R$, the supersymmetry transformation 
parameter must obey the condition Eq.~(\ref{eq:xi-cond}).
On the other hand, the boundary conditions for the fields, 
Eqs.~(\ref{eq:hyperparity},~\ref{eq:gaugeparity}) implies that 
the supersymmetry transformation parameter must obey
\begin{equation}
  \xi^i(y)  = U^i_j (\sigma_3)^j_k \gamma_5 \xi^k(-y), \qquad
  \xi^i(y') = U^{\prime i}_j (\sigma_3)^j_k \gamma_5 \xi^k(-y'),
\label{eq:orbixi}
\end{equation}
to preserve the boundary conditions of the fields. The number of 
supersymmetries on the boundaries is then determined by these two 
conditions, Eq.~(\ref{eq:xi-cond},~\ref{eq:orbixi}).

Let us focus on the $y=0$ boundary (the discussion for the $y=\pi R$ 
boundary is identical).  We first consider the case $\alpha_1 
= \alpha_2 = 0$.  In this case, Eq.~(\ref{eq:xi-cond}) and 
Eq.~(\ref{eq:orbixi}) become identical; in other words, the Killing 
spinor equation does not give an additional constraint on the 
transformation parameter $\xi^i$ beyond the one arising from the 
boundary conditions, Eq.~(\ref{eq:orbixi}).  This situation 
is similar to the $SU(5)$ example discussed in 
sub-section~\ref{subsec:type1-defect}.  In fact, we find 
that the $y=0$ brane is supersymmetry breaking defect of 
type~I, on which the 4D $N=2$ supersymmetry in the bulk is 
broken to 4D $N=1$. The number of supersymmetries can easily be 
understood from Eq.~(\ref{eq:orbixi}): $\xi^i(y) = (\sigma_3)^i_j 
\gamma_5 \xi^j(-y)$ requires half of $\xi^i$ to vanish at $y=0$. 
A defect of type~I implies that the symmetries can be consistently 
gauged, {\it i.e.} the theory can be embedded into supergravity. 
In supergravity, the gravitino $\psi_{3/2}$ obeys the boundary 
conditions analogous to Eq.~(\ref{eq:orbixi}). Thus, when expanded 
into 4D modes, the number of $\psi_{3/2}$'s and the number of 
$\xi$'s are the same (there is no need to impose any extra constraint 
on the gravitino field), ensuring the consistency of the theory with 
local supersymmetry.  In particular, if we choose $\alpha_1 = \alpha_2 
= \alpha'_1 = \alpha'_2 = 0$, the resulting theory possesses unbroken 
4D $N=1$ supersymmetry, whose transformation parameter is given 
by $\xi^1(x^\mu,y) = \exp(-\sigma/2)\xi_L$ where $\xi_L$ is 
a spinor (dependent on coordinates in supergravity) subject to 
the condition $\gamma_5 \xi_L = \xi_L$.  The explicit realization 
of this case in the context of supergravity has been extensively 
studied~\cite{Altendorfer:2000rr}.

We next consider the case where either $\alpha_1$ or $\alpha_2$ is 
non-zero. In this case Eq.~(\ref{eq:xi-cond}) and Eq.~(\ref{eq:orbixi}) 
give different conditions, and we find that the solution to both 
equations is only the trivial one, $\xi^i = 0$.  This implies that 
we do not have any supersymmetry on the $y=0$ brane.  Since the 
constraint $\xi^i(y=0) = 0$ is an extra condition imposed on $\xi$, 
additional to the one arising from the boundary conditions, 
the situation is similar to the $U(1)$ example discussed in 
sub-section~\ref{subsec:type2-defect} with $\xi^i(y=0)=0$ 
corresponding to Eq.~(\ref{eq:u1-gtp-const}). The defect is type~II 
and does not allow gauging of the supersymmetry of the theory.
The argument is similar to the previous $U(1)$ case. When we gauge 
supersymmetry, we must introduce the gravitino field and impose 
boundary conditions like Eqs.~(\ref{eq:orbixi}). Since the 
gravitino is a dynamical field, we cannot impose any additional 
constraint by hand.  This implies that the number of $\xi$'s is one 
smaller than that of $\psi_{3/2}$'s (in the 4D picture) due to 
the extra constraint $\xi^i(y=0)=0$. Since the consistent treatment 
of a spin-$3/2$ field requires a supersymmetry, this leads to 
an inconsistency; for instance, in such theories ghosts can be 
physically produced and certain scattering amplitudes lead to negative 
probabilities (the presence of such ghosts was also noted in 
Ref.~\cite{Gherghetta:2002nr}).

Why do we insist on gauging supersymmetry?  If supersymmetry were 
not a spacetime symmetry, we would be able to consider only global 
supersymmetry. We would be able to use arbitrary values for 
$\alpha_{1,2}$ and $\alpha'_{1,2}$ to construct models, in 
which supersymmetry is broken by boundary conditions.  However, 
supersymmetry {\it is} spacetime symmetry.  When we include gravity, 
we have to consider supergravity, in which supersymmetry is gauged.
This means that the boundaries at $y=0$ and $\pi R$ must be symmetry 
breaking defects of type~I: $\alpha_{1,2}$ and $\alpha'_{1,2}$ must 
be zero.  Therefore, we arrive at the following conclusion. 
{\it In AdS$_5$ the compactification on $S^1/Z_2$ is unique: 
we cannot use boundary conditions to break all bulk supersymmetries 
in a warped extra dimension}.

Nevertheless, in the next section we consider models on the truncated 
AdS$_5$ in which supersymmetry is broken by boundary conditions. 
As mentioned at the end of sub-section~\ref{subsec:type2-defect}, 
we do this because some theories can mimic certain properties of the 
theory with boundary condition supersymmetry breaking.  For instance, 
consider a theory with $\alpha_1 = \alpha_2 = \alpha'_1 = \alpha'_2 
= 0$ and break supersymmetry spontaneously by the expectation value 
for the $F$-component of a brane-localized chiral superfield $Z$ 
at $y=\pi R$. Then, if this expectation value is large (we can 
formally take $F_Z \rightarrow \infty$), we find that some properties 
of the boundary condition breaking, such as strict relations between 
supersymmetry breaking masses and the KK mass scale, are 
recovered~\cite{Arkani-Hamed:2001mi}.\footnote{
Maintaining the background geometry with a non-vanishing $F_Z$ may 
require certain compensating terms on boundaries, but we assume 
these terms do not affect the spectrum of the theory significantly.}
Thus, although the models presented in section~\ref{sec:model} do 
not allow consistent inclusion of gravity as they are, we think 
that it is worthwhile constructing some representative models and 
exploring their phenomenology.

We note that the case of supersymmetry breaking by boundary conditions
in flat space is now very simple to analyze. The Killing spinor
equations become trivial, with $\xi^i$ becoming constant for a global
transformation; crucially there is no Killing spinor constraint at the
boundary, such as Eq.~(\ref{eq:xi-cond}).  Therefore, the issue we have 
in AdS space, {\it i.e.} the incompatibility of Eq.~(\ref{eq:orbixi}) 
with Eq.~(\ref{eq:xi-cond}), does not exist in flat space.  Thus, 
any choices for the matrices $U^i_j$ and $U^{\prime i}_j$ in 
Eq.~(\ref{eq:orbixi}) yield type~I defects at both boundaries, where 
each boundary preserves a single 4D supersymmetry, with the orientation 
of the supersymmetry in $SU(2)_R$ space depending on the parameter 
$\alpha$ relevant at that boundary. The entire system preserves 
a supersymmetry in 4D only if the two boundaries preserve the same 
supersymmetry, $\alpha = \alpha'$, otherwise supersymmetry is completely 
broken by the boundary conditions. If either boundary is allowed to 
have a Lagrangian which is not invariant under any supersymmetry, the 
resulting $N=0$ defect is of type~II, so that the resulting theories 
are inconsistent with supergravity. 

Finally in this section, we complete the Lagrangian in the case of 
$\alpha_1 = \alpha_2 = \alpha'_1 = \alpha'_2 = 0$.  The bulk Lagrangian 
of Eqs.~(\ref{eq:hyperaction},~\ref{eq:hypermass},~\ref{eq:gaugelag}) 
is not invariant under the supersymmetry transformation at $y=0$ 
and $\pi R$.  For instance, when we vary the hypermultiplet action 
Eqs.~(\ref{eq:def-s},~\ref{eq:hyperaction}), we find that the terms 
that spoil invariance appear from $\partial_y$ acting on $\sigma'$:
\begin{equation}
  \sqrt{-g}\delta{\cal L}_{\rm hyp.kin.} 
    = \sqrt{-g}\left[\cdots 
    + \frac32 \sigma''\sqrt2 \bar{\Psi}\xi^i\phi_i + \mbox{h.c.} \right],
\end{equation}
where $\sigma'' = 2k(\cdots+\delta(y)-\delta(y-\pi R)+\cdots)$. 
However, these terms can be canceled if we add brane mass terms 
for the scalars:
\begin{equation}
  {\cal L}_{\rm hyp.kin.} \rightarrow 
  {\cal L}_{\rm hyp.kin.} - \frac32 \sigma'' \phi^*_i \phi^i.
\end{equation}
This gives the correct supersymmetric Lagrangian on AdS$_5$ 
compactified on $S^1/Z_2$. A similar analysis for 
Eqs.~(\ref{eq:hypermass},~\ref{eq:gaugelag}) leads to 
\begin{eqnarray}
  && {\cal L}_{\rm hyp.mass} \rightarrow 
  {\cal L}_{\rm hyp.mass} + c \sigma'' (\sigma_3)^i_j \phi^*_i \phi^j,
\\
  && {\cal L}_{\rm gauge} \rightarrow
  {\cal L}_{\rm gauge} - \frac{1}{g^2} \sigma'' \Sigma^2.
\end{eqnarray}
After integrating out the auxiliary fields, these Lagrangians agree 
with the on-shell Lagrangian given in Ref.~\cite{Gherghetta:2000qt}.

\section{Models}
\label{sec:model}

\subsection{Warped constrained standard model}
\label{subsec:wcsm}

Consider an $SU(3)_C \times SU(2)_L \times U(1)_Y$ supersymmetric gauge
theory on truncated AdS$_5$ space. Each 4D boundary is necessarily a
defect in the space of supersymmetries, since the two bulk supersymmetries
cannot coexist on a 4D boundary. In the last section we have shown that 
if both defects are of type~I, then the supersymmetries preserved at 
each boundary must align with each other, so that the entire system
preserves a 4D supersymmetry.  To break supersymmetry by boundary
conditions, we must consider supersymmetry breaking by means of 
a defect of type~II.  Furthermore, we assume that the Planck brane 
located at $y=0$ is a type~I defect preserving one supersymmetry, since, 
if it were type~II, all supersymmetries would be broken at the Planck 
scale.\footnote{
We could instead choose the Planck brane to be a type~II defect, if 
we localize the Higgs fields to the TeV brane. Such a construction 
can lead to theories where there is a little hierarchy between the 
electroweak and new physics scales~\cite{Gherghetta:2003wm}.}
Therefore the TeV brane at $y=\pi R$ must be of type~II, so that 
couplings on this brane explicitly break all supersymmetries.

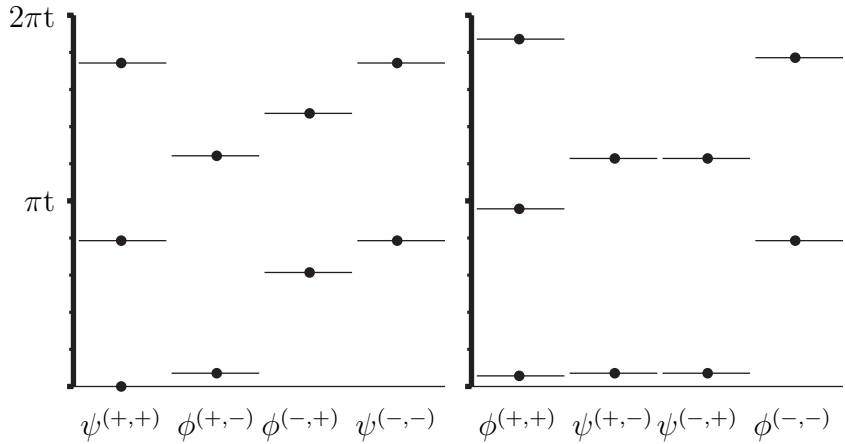
\begin{figure}
\begin{center}
\begin{picture}(300,150)(0,0)
 \SetOffset(0,5)
 \LinAxis(10,10)(10,150)(2,5,2,0,2)
 \Line(10,10)(150,10)
 \Vertex(28,10){2}
 \Line(12,65)(45,65) \Vertex(28,65){2}
 \Line(117,65)(150,65) \Vertex(132,65){2}
 \Line(12,132)(45,132) \Vertex(28,132){2}
 \Line(117,132)(150,132) \Vertex(132,132){2}
 \Line(82,53)(115,53) \Vertex(99,53){2}
 \Line(82,113)(115,113) \Vertex(99,113){2}
 \Line(47,15)(80,15) \Vertex(64,15){2}
 \Line(47,97)(80,97) \Vertex(64,97){2}
 \Text(28,2)[t]{$\psi^{(+,+)}$}
 \Text(64,2)[t]{$\phi^{(+,-)}$}
 \Text(96,2)[t]{$\phi^{(-,+)}$}
 \Text(132,2)[t]{$\psi^{(-,-)}$}
 \Text(3,80)[r]{$\pi$t}
 \Text(3,150)[r]{$2\pi$t}
 \LinAxis(160,10)(160,150)(2,5,2,0,2)
 \Line(160,10)(300,10)
 \Line(162,14)(195,14) \Vertex(178,14){2}
 \Line(162,77)(195,77) \Vertex(178,77){2}
 \Line(267,65)(300,65) \Vertex(282,65){2}
 \Line(162,141)(195,141) \Vertex(178,141){2}
 \Line(267,134)(300,134) \Vertex(282,134){2}
 \Line(232,15)(265,15) \Vertex(249,15){2}
 \Line(232,96)(265,96) \Vertex(249,96){2}
 \Line(197,15)(230,15) \Vertex(214,15){2}
 \Line(197,96)(230,96) \Vertex(214,96){2}
 \Text(178,2)[t]{$\phi^{(+,+)}$}
 \Text(214,2)[t]{$\psi^{(+,-)}$}
 \Text(246,2)[t]{$\psi^{(-,+)}$}
 \Text(282,2)[t]{$\phi^{(-,-)}$}
\end{picture}
\caption{Mass spectrum for matter-like fields (left) with $c_M=1/2$ 
 and for Higgs-like fields (right) with $c=1/2$ and $r'=-1$.}
\label{fig:spectrum}
\end{center}
\end{figure}

The field content and boundary conditions are chosen to be identical 
to those of the constrained standard model~\cite{Barbieri:2000vh}, 
so that gauge, matter, and a single Higgs hypermultiplet all propagate 
in the bulk. Since there is a single zero-mode Higgs boson, we 
expect a Higgs sector far more constrained than that of the MSSM. 
The supersymmetry breaking boundary conditions are those of 
Eqs.~(\ref{eq:hyperparity},~\ref{eq:gaugeparity}) with $\alpha'_2=1/2$, 
$\alpha_{1,2} = \alpha'_1 = 0$ and $P_{\rm matter}=P'_{\rm matter}=+1$, 
$P_{\rm Higgs}=-P'_{\rm Higgs}=+1$.  The mass spectrum for both 
matter-like and Higgs-like boundary conditions are shown in 
Fig.~\ref{fig:spectrum}. While these boundary conditions are identical 
to those of the flat space constrained standard model, we stress 
that in that theory both boundary defects were of type~I preserving 
orthogonal supersymmetries, so that the structure of supersymmetry 
breaking differs greatly in the warped case. We assume that all matter 
hypermultiplets have a bulk mass $c_M=1/2$, ensuring that the quark 
and lepton zero modes are conformally flat. This is analogous to the 
flat space theory in the absence of bulk masses. We expect that 
deviations from $c_M= 1/2$ would be analogous to introducing bulk 
masses in the flat case~\cite{Marti:2002ar}. To obtain a predictive 
theory of electroweak symmetry breaking with a single Higgs boson, 
the Higgs must propagate in the bulk. If we had instead placed the 
Higgs boson on the Planck brane, then 4D supersymmetry on that brane 
would have prevented it from generating down-type masses.  If we had 
placed it on the non-supersymmetric TeV brane, the quartic coupling 
would be arbitrary and there would be no prediction for the physical 
Higgs boson mass.  A bulk Higgs boson, however, is able to generate 
up-type masses at the $y=0$ brane and all masses at the $y=\pi R$ 
brane, and to a large extent radiative corrections are controlled 
by the unbroken bulk supersymmetry.

The bulk mass for the Higgs hypermultiplet, $c$, is still free as 
is the Higgs boson brane mass term at the TeV brane ${\cal L} 
= r'(\phi^{1*}\phi^{1}) 2k\delta(y-\pi R)$, where $\phi^{1}$ is 
the first component of a complex $\mathrm{SU}(2)_R$ doublet and gauge 
indices are contracted.  These two parameters in turn determine the 
profile of the Higgs boson KK modes in the bulk, and the tree-level 
mass of the Higgs boson KK modes.  In particular the lightest mode 
mass is approximately given by:
\begin{equation}
  m_{\rm tree} 
    \simeq 2 \sqrt{(c^2-1/4)\frac{3/2-r'-c}{5/2-r'+c}}
      \left( \frac{t}{k}\right) ^{c-1/2}t,
\end{equation}
for $(t/k)^{c-1/2} \ll 1$, where $t=e^{-k\pi R}k$ is the scale of 
physics at the $y=\pi R$ brane. An important result is that for 
$c>1/2$ the tree-level mass is much smaller than the typical KK mass 
scale $t$ so that the full Higgs mass parameter becomes only weakly 
sensitive to $r'$. This is because the lowest level wavefunction is 
strongly peaked around the $y=0$ brane. Therefore, while we have 
no knowledge of the TeV brane parameters (and radiative corrections 
to some of these parameters are even power divergent), the low energy 
physics is largely insensitive to their values.

\begin{figure}
\begin{center}
  \includegraphics[width=10cm]{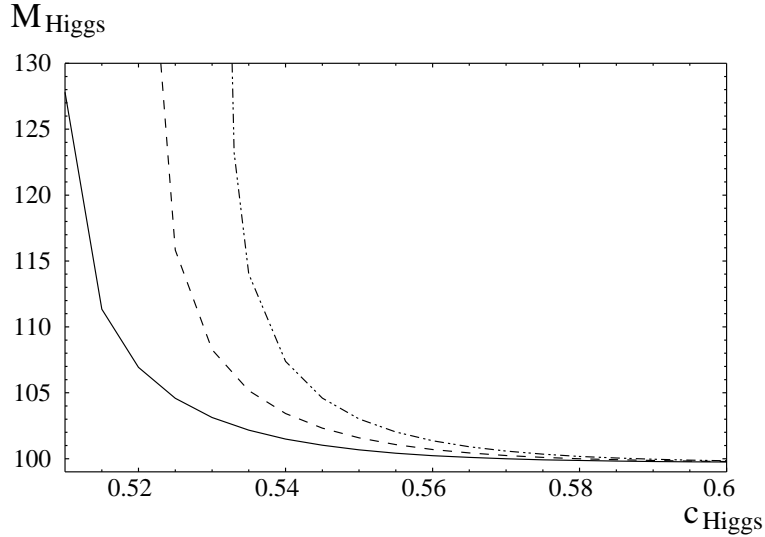}
\caption{Physical Higgs boson mass in GeV for $r'=-1$ (dash-dot-dot), 
 $r'=0$ (dashed) and $r'=0.5$ (solid).}
\label{fig:higgsmass}
\end{center}
\end{figure}

Since the tree level mass of the Higgs boson rapidly becomes 
small for $c>1/2$, electroweak symmetry breaking is triggered 
radiatively via the top Yukawa coupling, which we assume to be 
located dominantly on the Planck brane.  As discussed in 
section~\ref{subsec:boundary}, the supercurrent is conserved 
locally in the bulk, so radiative effects must respect supersymmetry 
there.  Therefore, supersymmetry guarantees that the bulk Higgs 
mass is not renormalized. Thus, as in models of boundary condition 
supersymmetry breaking on flat extra dimension, we expect corrections 
to the 4D Higgs boson mass to be finite, except for the contribution 
from the $y=\pi R$ brane, which we have argued is small. Therefore, 
by taking $r'$ to be the renormalized brane mass we are able to 
calculate the physical Higgs mass in terms of $c$ and $r'$. 
We have computed radiative corrections from the top quark Yukawa 
coupling to the Higgs boson effective potential. After minimizing 
this effective potential, the mass scale of the KK modes is 
determined from $M_Z$, and the resulting prediction for the Higgs 
boson mass is shown in Fig.~\ref{fig:higgsmass}, for a range of 
$c$ and for three values of $r'$.  The insensitivity to $r'$ as $c$ 
increases above $1/2$ is striking, but not unexpected as in this 
region the tree-level mass effectively vanishes.  For the same reason, 
the physical Higgs boson mass becomes constant for large $c$ at about 
$100~{\rm GeV}$.  One might worry that direct searches rule out 
a single Higgs boson with mass less than $115~{\rm GeV}$, and the 
model requires a large degree of fine tuning to reach such a mass. 
However, since $r'$ is not the only TeV brane operator that affects 
the Higgs mass, we expect that there are $O(15\%)$ corrections 
to our calculation.  For example, there can be additional quartic 
interactions, top Yukawa couplings and terms involving the $F$-fields 
of the matter multiplets, on the TeV brane. Although all of these 
terms are suppressed by the wavefunction overlaps of the various 
fields with the TeV brane and thus introduce only small corrections, 
they can give non-negligible effects on the physical Higgs boson mass; 
for example, a brane-localized quartic coupling is expected to 
introduce a $\simlt 15\%$ correction in the physical Higgs mass in the 
limit of strong coupling. Therefore, taking note of these possible 
corrections, the model is not ruled out for a reasonably wide range 
of parameter space. Notice that for a given $r'$, electroweak symmetry 
is not broken for all $c$.  Below a certain $c$, the radiative 
corrections are unable to overcome the positive tree level mass
squared and electroweak symmetry breaking does not occur. Thus the 
curves in Fig.~\ref{fig:higgsmass} end at these points.
\begin{figure}[tbh]
\begin{center}
  \includegraphics[width=10cm]{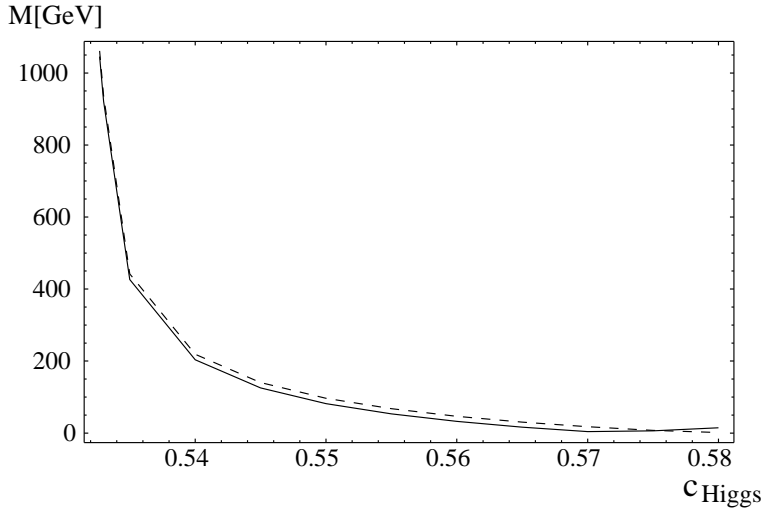}
\caption{Lightest neutralino (solid) and lightest chargino (dashed) 
 masses in GeV for $r'=-1$ and $z=-0.75$.}
\label{fig:nlsp}
\end{center}
\end{figure}
Taking $c>1/2$ also leads to light Higgsinos. The lightest Higgsino 
mass is approximately
\begin{equation}
  m_{\tilde{h}} \simeq 2\sqrt{c^2 -1/4}\left( \frac{t}{k}\right) ^{c-1/2}t.
\end{equation}
Notice that no brane mass term can be written for the Higgsinos because
$\tilde{h}^{1}$ vanishes at $y=\pi R$.  The masses of the lightest
neutralino and lightest chargino therefore place a bound on $c$. 
However, this bound is weak because it depends on the size of the 
brane-localized kinetic term for $\tilde{h}^{2}$
\begin{equation}
  S_{4} = -\int d^{4}x
    (k/t)^{-1}Z\bar{\tilde{h}}^{2}\partial_{\mu}\hat{\gamma}^{\mu}
    \tilde{h}^{2}\left( \eta_{\tilde{h}^{2}}(\pi R) \right)^2,
\end{equation}
where indices are raised and lowered with $\eta^{\mu\nu}$ and 
$\hat{\gamma}^{\mu}$ are the four dimensional Dirac gamma matrices. 
$\eta_{\tilde{h}^{2}}(y)$ is the wavefunction of the lightest 
right-handed Higgsino. If we consider the dimensionless combination 
$z=Z(k/t)^{-1}\left( \eta_{\tilde{h}^{2}}(\pi R)\right) ^2$ we find 
that the four dimensional kinetic term for the lightest $\tilde{h}^{2}$ 
has a coefficient $\simeq 1+z$ and that going to canonical normalization 
the lightest Higgsino mass is
\begin{equation}
  m_{\rm canonical} \simeq \frac{m_{\tilde{h}}}{\sqrt{1+z}}.
\end{equation}
Notice that the strong peaking of the wavefunction enhances the effect 
due to the brane kinetic term, so that a correction to the Higgsino 
mass of order unity is expected.  As an example, we show in 
Fig.~\ref{fig:nlsp} the lightest chargino and neutralino for the 
case that $z=-0.75$ and $r'=-1$, though there is only weak sensitivity 
to $r'$.  In this case, the chargino mass is not ruled out by direct 
searches for $c \simlt 0.55$. 

By introducing a type~II supersymmetry breaking defect on a warped 
background, we are able to construct a predictive theory of electroweak 
symmetry breaking with one Higgs doublet. The theory requires 
a moderate peaking of the Higgs boson on the $y=0$ brane, which could 
be the origin of the $m_t/m_b$ ratio. While the Higgs boson is expected 
to be close to its experimental limit of $115~{\rm GeV}$, a precise 
prediction is not possible because of the degree of peaking of the 
Higgs boson wavefunction and the supersymmetry breaking interactions 
on the TeV brane. The lightest chargino and neutralino are also close 
to their experimental bounds.

\subsection{Twisted warped grand unified theory}
\newcommand{\tw}{({\bf 1},{\bf 2})_{\frac12~}}
\newcommand{\twbar}{({\bf 1},{\bf 2})_{-\frac12}}
\newcommand{\thh}{({\bf 3},{\bf 1})_{-\frac13}}
\newcommand{\thbar}{(\bar{\bf 3},{\bf 1})_{\frac13~}}
\newcommand{\tq}{({\bf 3},{\bf 2})_{\frac16~}}
\newcommand{\tqbar}{({\bar{\bf 3}},{\bf 2})_{-\frac16}}

While the theory just described has logarithmic running of gauge
couplings up to the mass scale of the Planck brane, it is not a
theory of gauge coupling unification, since nothing in the theory
requires the bulk gauge couplings to be unified.  In order to
construct such a theory we must consider a model in which the bulk
Lagrangian is symmetric under some grand unified group.  Here we 
consider an $SU(5)$ supersymmetric gauge theory on a slice of 
$\mathrm{AdS}_5$.  We take the warped supersymmetric grand unified 
theory of Ref.~\cite{Goldberger:2002pc}, with all matter and Higgs 
in the bulk, and break the $SU(5)$ symmetry by boundary conditions 
imposed at the Planck brane.  Each generation contains 
${\cal F}: \{ F(\bar{\bf 5}), F^c({\bf 5}) \} + {\cal F}': 
\{ F'(\bar{\bf 5}), F^{\prime c}({\bf 5})\}$ and 
${\cal T}: \{ T({\bf 10}), T^c(\bar{\bf 10})\} + {\cal T}': 
\{ T'({\bf 10}), T^{\prime c}(\bar{\bf 10})\}$ where 
$\Phi(R)$ represents a chiral supermultiplet in the $R$ 
representation of $SU(5)$. There are, in addition, two Higgs 
hypermultiplets ${\cal H}: \{ H({\bf 5}), H^c(\bar{\bf 5})\}$ and 
$\bar{\cal H}: \{ \bar{H}(\bar{\bf 5}), \bar{H}^{c}({\bf 5})\}$.
In this model, the boundary conditions are given such that each brane 
is a symmetry breaking defect of type~I with respect to supersymmetry.
The Planck brane is additionally a symmetry breaking defect of type~I
with respect to the gauge group, breaking $SU(5) \rightarrow SU(3)_C
\times SU(2)_L \times U(1)_Y$. The TeV brane respects the full $SU(5)$
group. (The boundary conditions for some of the bulk fields are 
given explicitly in Table~\ref{fig:notwist}.)  The zero-mode particle 
content of the model is the same as in the MSSM, while the KK towers 
consist of $SU(5)$ symmetric particles of masses around TeV.  It was 
shown in~\cite{Goldberger:2002pc} that, if all bulk fields carry 
$c \geq 1/2$, only the zero modes contribute to the differential 
gauge coupling running, so that the model leads to the same beta 
functions as in the MSSM.  Therefore, despite the drastic departure 
from the MSSM particle content at the TeV scale, the theory preserves 
logarithmic gauge coupling unification at a high scale.
\begin{table}
\begin{center}
\begin{tabular}{|lcc|lcc|} \hline
  \multicolumn{3}{|c|}{Matter} & \multicolumn{3}{c|}{Higgs} \\
  \cline{1-6} 
    &  $F_{++}\twbar$  &  $F_{-+}\thbar$  &
    &  $H_{++}\tw$     &  $H_{-+}\thh$    \\ 
  $\mathcal{F}'$  &  $F^c_{--}\tw$     &  $F^c_{+-}\thh$ & 
  $\mathcal{H}$   &  $H^c_{--}\twbar$  &  $H^c_{+-}\thbar$
    \\ \cline{1-6} 
    &  $F_{-+}\twbar$  &  $F_{++}\thbar$  &
    &  $H_{++}\twbar$  &  $H_{-+}\thbar$  \\ 
  $\mathcal{F}$        &  $F^c_{+-}\tw$  &  $F^c_{--}\thh$  &
  $\bar{\mathcal{H}}$  &  $H^c_{--}\tw$  &  $H^c_{+-}\thh$  \\ \hline
\end{tabular}
\caption{Superfields from matter and Higgs fields are listed by 
 their quantum numbers and parity assignments before the supersymmetry 
 breaking twist.  The quantum numbers represent those under 
 $SU(3)_C \times SU(2)_L \times U(1)_Y$. The far left column 
 indicates what hypermultiplet the fields are contained in. 
 $\mathcal{F}$ and $\mathcal{F}'$ are $\bar{\bf 5}$'s of matter 
 while $\mathcal{H}$ and $\bar{\mathcal{H}}$ are Higgs multiplets.}
\label{fig:notwist}
\end{center}
\end{table}

With the boundary conditions described above, both branes are defects 
of type~I respecting the same 4D $N=1$ supersymmetry, so that there 
exists unbroken $N=1$ supersymmetry in the low-energy 4D theory. 
One way of breaking this remaining supersymmetry is to consider 
a supersymmetry breaking VEV located on the TeV brane. Instead, here 
we consider breaking the remaining supersymmetry by modifying the 
boundary conditions such that the brane at $y=\pi R$ becomes a 
supersymmetry breaking defect of type~II, as in the previous 
sub-section.  This is accomplished by introducing non-zero $\alpha'$ 
parameters in Eqs.~(\ref{eq:hyperparity},~\ref{eq:gaugeparity}). 
Without loss of generality, we can take $\alpha'_1=\alpha_{1,2}=0$.
We here choose the supersymmetry breaking parameter $\alpha'_2=1/2$.
We also choose $P'_{\rm Higgs}=-1$ so that there are light Higgs 
scalars from the Higgs hypermultiplets.

\begin{table}
\begin{center}
\begin{tabular}{|lcc|lcc|} \hline
  \multicolumn{3}{|c|}{Matter} & \multicolumn{3}{c|}{Higgs} \\
    \cline{1-6} 
  & $\phi_{+-}\twbar$    & $\phi_{--}\thbar$ &
  & $\phi_{++}\tw$       & $\phi_{-+}\thh$    \\ 
  $\mathcal{F}'$   & $\psi_{++}\twbar$ & $\psi_{-+}\thbar$ & 
  $\mathcal{H}$          & $\psi_{+-}\tw$    & $\psi_{--}\thh$ \\
  & $\phi^c_{-+}\tw$   &
  $\phi^c_{++}\thh$    & & $\phi^c_{--}\twbar$ &
  $\phi^c_{+-}\thbar$ \\ 
  & $\psi^c_{--}\tw$   &
  $\psi^c_{+-}\thh$    & & $\psi^c_{-+}\twbar$ &
  $\psi^c_{++}\thbar$ \\  \cline{1-6} 
  & $\phi_{--}\twbar$  & $\phi_{+-}\thbar$     &
  & $\phi_{++}\twbar$  & $\phi_{-+}\thbar$     \\ 
  $\mathcal{F}$        & $\psi_{-+}\twbar$     &  $\psi_{++}\thbar$&
  $\bar{\mathcal{H}}$  & $\psi_{+-}\twbar$     &  $\psi_{--}\thbar$ \\ 
  & $\phi^c_{++}\tw$   &
  $\phi^c_{-+}\thh$    & & $\phi^c_{--}\tw$  & 
  $\phi^c_{+-}\thh$  \\ 
  & $\psi^c_{+-}\tw$   & $\psi^c_{--}\thh$   &
  & $\psi^c_{-+}\tw$   & $\psi^c_{++}\thh$ \\ \hline
\end{tabular}
\caption{Fields from $\bar{\bf 5}$ matter and Higgs multiplets are 
 listed by their quantum numbers and parity assignments after the 
 supersymmetry breaking twist. $\phi$ and $\phi^c$ ($\psi$ and $\psi^c$) 
 represent complex scalar (Weyl fermion) fields in $\Phi$ and $\Phi^c$ 
 superfields, respectively, where $\Phi = F, F', H, \bar{H}$.}
\label{fig:twist}
\end{center}
\end{table}
We first consider the effect of the supersymmetry breaking twist, 
$\alpha'_2 \neq 0$, on the matter multiplets.  Here, the net effect 
is that the parity under $y' \rightarrow -y'$, $Z'$, changes sign 
for the $SU(2)_R$ doublet scalars. As a consequence, the MSSM sfermions 
no longer possess a zero mode while their first KK modes appear at 
$O(\mathrm{TeV})$.  However, a new scalar zero mode now appears. 
As shown in Tables~\ref{fig:twist} and \ref{fig:zeromodes}, these 
scalars are related to the standard model fermions by the broken 
generators of $SU(5)$ and by the supersymmetry broken at the Planck 
brane.  We will call these the $SU(5)$, $N=2$ partners of the standard 
model fermions.  Notice that a full generation will possess both a 
fermion and a scalar with conjugate quantum numbers.  Therefore, 
mere observation of quantum numbers and the mass spectrum (before 
electroweak symmetry breaking) could mimic the presence of an unbroken 
supersymmetry.  However, the fermion and scalar originate from 
different hypermultiplets and thus there is no supersymmetry that 
relates the two.  For example, the two fields will not be coupled 
by the gauginos.  The extra bosonic fields can be made heavy by mass 
terms located on the $y=\pi R$ brane.  In order for these fields 
to become sufficiently heavy their wavefunctions must have 
a sizable overlap with the TeV brane.  This translates into 
the requirement that $c_{\rm matter} \geq 1/2$, in accordance 
with the requirement for gauge coupling unification and stability 
of the proton~\cite{Goldberger:2002pc}.  The resulting masses are 
naturally in the TeV region.\footnote{
The extra scalars can also obtain masses through radiative corrections 
from gaugino masses. In the case where these masses are sufficiently 
large, we do not necessarily satisfy the conditions $c_{\rm matter} 
\geq 1/2$ to make these fields heavy, and the scalar mass squareds 
are one-loop smaller than the supersymmetry breaking scale $\sim t^2$.}

In the Higgs sector, because of the change in sign of $P'_{\rm Higgs}$ 
in addition to the supersymmetry breaking twist, the $Z'$ parities of 
the fermions change sign.  As a result, the doublet Higgsino becomes 
heavy. However, as in the matter sector, the $SU(5)$, $N=2$ partners 
of the two Higgs bosons, a pair of color triplet fermions, now possess 
zero modes. Again, as in the matter sector, these unwanted fields can be 
made massive via a TeV brane localized mass term: a Dirac mass for the 
two colored Higgsinos. Again, to give sufficient mass for the undesired 
states, the wavefunction overlaps of these states with the TeV brane 
must be sizable, requiring $c_{\rm Higgs} \geq 1/2$. Note that larger 
values of $c_{\rm Higgs}$, $c_{\rm Higgs} > 1/2$, have the added 
benefit of minimizing the influence of the supersymmetry breaking 
brane on the Higgs bosons as discussed in the previous sub-section. 
We expect electroweak symmetry breaking to proceed much like in the
previous sub-section including the issue of too light Higgsino 
doublets. To illustrate the effects of the supersymmetry breaking
boundary conditions, both the zero and non-zero modes of the $\bf 5$'s 
and $\bar{\bf 5}$'s of the $F$,$F'$ matter and Higgs fields are 
shown in Tables~\ref{fig:notwist} and \ref{fig:twist} both before 
and after the supersymmetry breaking twist.

We finally consider the gauge sector. This sector has a quite similar 
structure to the Higgs sector. The boundary condition twist at $y=\pi R$, 
$\alpha'_2=1/2$, acts on the $SU(2)_R$ doublets, so the $Z'$ parities 
for the gauginos change sign. The MSSM gauginos become heavy while the 
$SU(5)$, $N=2$ partners of the gauge bosons are made light by the 
supersymmetry breaking twist. However, these undesired fields can 
gain masses via a mass term on the TeV brane.  In the case of the 
gauge multiplets the bulk mass is required to be $c_{\rm gauge}=1/2$ 
by gauge invariance, and thus the zero-mode fermions have conformally 
flat wavefunctions, insuring that they have sizable wavefunction 
overlaps with the TeV brane.
\begin{table}
\begin{center}
\begin{tabular}{|cccccc|} \hline
  Standard Model Matter & $\psi\tq$ 
  & $\psi(\bar{\bf 3},{\bf 1})_{-\frac23}$ 
  & $\psi(\bar{\bf 3},{\bf 1})_{\frac13~}$ 
  & $\psi({\bf 1},{\bf 2})_{-\frac12}$ 
  & $\psi({\bf 1},{\bf 1})_{1}$
  \\ \cline{1-6}
  $SU(5), N=2$ Partners 
  & $\phi^{c}\tqbar$ 
  & $\phi^{c}({\bf 3},{\bf 1})_{\frac23~}$ 
  & $\phi^{c}({\bf 3},{\bf 1})_{-\frac13}$ 
  & $\phi^{c}({\bf 1},{\bf 2})_{\frac12~}$ 
  & $\phi^{c}({\bf 1},{\bf 1})_{-1}$
  \\ \hline
\end{tabular}
\caption{Matter fields which have zero modes: standard model quarks 
 and leptons and their $SU(5)$, $N=2$ partners.}
\label{fig:zeromodes}
\end{center}
\end{table}

Now, we consider the effect of supersymmetry breaking twist, 
$\alpha'_2 \neq 0$, on the gauge coupling unification. A theory 
on the truncated AdS$_5$ space has a different description in terms 
of a 4D quasi-conformal field theory~\cite{Arkani-Hamed:2000ds}. 
In this dual 4D picture, changing the boundary conditions at the 
TeV brane corresponds to changing the TeV physics, so that it can 
be viewed as an IR effect.  Therefore, the boundary condition breaking 
twist at the TeV brane is expected not to change the differential 
running above the TeV scale, and the model is expected to preserve the 
successful prediction for $\sin^2\theta_w$~\cite{Goldberger:2002pc}. 
This expectation can be confirmed by direct calculation of the beta 
functions with the twists, $\alpha'_2 = 1/2$ and $P'_{\rm Higgs} = -1$, 
using the formulae found in Ref.~\cite{Choi:2002ps}. Notice, however, 
that there is no energy range in which our theory mimics the MSSM 
particle content. If the brane mass terms are smaller than the 
KK mass gap, then the first new particles to be created will 
include the $SU(5)$, $N=2$ partners of the standard model particles. 
These include colored Higgsinos and gauginos of the broken $SU(5)$ 
generators. If the brane mass terms are larger than the KK mass gap, 
then the first new particles to be created will be the first KK mode 
which contains many states in addition to those of the MSSM. The model 
makes the same prediction for $\sin^2 \theta_w$ as the MSSM, despite 
these drastic departures from the MSSM particle content at the TeV scale.

We may also wish to maintain the feature of gauge coupling unification 
in the context of a theory with one Higgs doublet. In this case, 
it is easiest to return to the situation before the supersymmetry 
breaking twist was made. We can then imagine removing the 
$\{ \bar{H}(\bar{\bf 5}), \bar{H}^c({\bf 5})\}$, which contained 
the $\bar{H}({\bf 1},{\bf 2})_{-1/2}$ Higgs doublet as a zero mode. 
This change in the zero-mode particle content will change the beta 
function; we therefore also remove the $\{ F(\bar{\bf 5}), 
F^{c}({\bf 5})\}$ that had contained the third generation 
$F(\bar{\bf 3},{\bf 1})_{1/3}$.  The sum of these two zero-mode fields 
contribute to the differential running of the gauge couplings the 
same as a single $\bar{\bf 5}$, and therefore their removal does 
not affect gauge unification.  We will not worry about the missing 
$b_{R}$ at this stage; it will reappear after we break the remaining 
$N=1$ supersymmetry.

Now let us consider the $\bf 5$'s and $\bar{\bf 5}$'s of the third 
generation matter and Higgs fields after breaking supersymmetry. 
Notice that the $SU(5), N=2$ partner of the remaining Higgs
boson ({\it i.e.} the $\psi^c_{++}$ component of the ${\cal H}$ 
hypermultiplet) is a fermion with the same quantum numbers as $b_{R}$. 
We may identify this field as the right-handed bottom quark thereby 
completing the third generation. This identification points out the
fact that there no longer exists any distinction between Higgs-like
and matter-like boundary conditions.  Instead, we have unified matter
and Higgs in the context of an $SU(5)$ model: from the 5D point of 
view the quantum numbers and boundary conditions for the ${\cal H}$ 
hypermultiplet and the two ${\cal F}$ hypermultiplets are the same.
There are three potential Higgs bosons, each one an $SU(5),N=2$ 
partner of a right-handed down-type quark.  However, since the 
${\cal H}$ hypermultiplet has the bulk mass parameter such that 
the zero-mode doublet scalar is localized toward the Planck brane 
while the zero-mode scalars from the ${\cal F}$ hypermultiplets 
are localized to the TeV brane, it is natural that only two of these 
scalars receive $O(\mathrm{TeV})$ masses from the $y=\pi R$ brane, 
leaving one light Higgs doublet. This light Higgs field will develop 
an electroweak breaking expectation value through radiative corrections, 
as in the previous models. Therefore, we can naturally obtain the 
theory with logarithmic gauge coupling unification, which effectively 
has only a single Higgs boson. Note that the Yukawa couplings for the 
up-type quarks can be located both on the Planck and TeV branes but 
those for the down-type quarks (and charged leptons) can be located 
only on the TeV brane.  Thus we naturally understand the origin of 
the $m_t/m_b$ ratio in this theory through a moderate peaking of the 
Higgs wavefunction toward the Planck brane.

\section{Conclusions}
\label{conclusion}

In this paper we have studied the properties of boundaries in higher 
dimensional theories that do not respect the full symmetries of 
the bulk.  Such defects break symmetries explicitly, but since all the 
breakings are localized on boundaries, local counterterms in the bulk 
are restricted in exactly the same manner as they would be in the 
absence of the defects.  In particular, the effects of explicit 
breaking are suppressed by the volume of the extra dimensions and/or 
small wavefunction overlaps in the low energy 4D theories. Therefore, 
symmetry breaking by point defects provides an interesting alternative 
to spontaneous symmetry breaking, in which we can systematically 
suppress the size of explicit breaking and can use the symmetry 
to control radiative corrections.

There are two different classes of symmetry breaking defects. 
A type~I defect possesses the maximum symmetry allowed by the boundary
conditions of the fields. Put another way, if the symmetry is made
local the boundary conditions of the gauge parameters and gauge fields
coincide.  The two therefore have the same KK decomposition, and there
exists a gauge transformation corresponding to each gauge field in the
4D theory. Such a theory can therefore be consistently gauged.  If 
a theory possesses a defect of type~II, then the symmetry may not be
gauged.  Type~II defects arise by requiring the brane Lagrangian to
be invariant under a smaller symmetry than that allowed by the
boundary conditions. In order to enforce this, one must impose
additional constraints on the symmetry transformation parameters at 
the boundaries. If one were to attempt to gauge such a theory, the
additional boundary conditions on the gauge parameters, as compared to
the gauge fields, would result in gauge fields without corresponding
gauge transformations.  This will therefore result in inconsistencies
such as states with negative norm. These results apply to any theory in 
which a symmetry is broken by boundary conditions on extra dimensions.

In particular, these ideas may be applied to supersymmetry 
breaking on a slice of $\mathrm{AdS}_5$.  In this case, the Killing
spinor equations result in a non-trivial constraint on the
supersymmetry transformation parameters at the boundaries of 
the space.  If only half of the supersymmetry is broken, reducing 
the four dimensional $N=2$ to $N=1$, then the Killing spinor
condition coincides with the conditions required by the boundary
conditions of the fields, and the boundaries become symmetry 
breaking defects of type~I.  However, if one attempts to break 
all of the supersymmetries, then the Killing spinor equations do 
in fact constitute an additional constraint and at least one boundary 
must be a type~II defect.  As a consequence, unlike flat space, 
breaking supersymmetry by boundary conditions on a warped background 
is not consistent with supergravity.

Despite this difficulty, we argue that a theory with supersymmetry
breaking boundary conditions may approximate a theory that is
consistent with gravity. We therefore presented two models that
make use of this mechanism. First, we constructed a theory of
electroweak symmetry breaking with a single Higgs doublet: a warped
version of the constrained standard model. Supersymmetry greatly
constrains the Higgs potential, while a bulk mass for the Higgs
hypermultiplet reduces the sensitivity to the dynamics of the
supersymmetry breaking brane.  As a result, we have succeeded in
constructing a predictive theory leading to a Higgs boson mass that 
may be close to its experimental lower bound.

In our second model we constructed a theory of gauge coupling
unification in which both the grand unified group and supersymmetry
are broken by boundary conditions on the same extra dimension. While
gauge coupling unification occurs as in the MSSM, the low energy
particle content may deviate drastically from that of the MSSM.
Depending on the mass parameters of the supersymmetry breaking brane,
it is possible that the lowest mass gaugino may be the super partner
of the broken gauge bosons. It is also possible that the lightest
Higgsinos are colored.  We have also shown that it is possible to 
remove the Higgs hypermultiplets from this model and identify the 
Higgs boson as one of the $SU(5), N=2$ partners of the right handed 
down-type quarks and in this way unify the Higgs and matter 
in the context of an $SU(5)$ grand unified theory.

\section*{Acknowledgements}
\label{acknowledgements}

The work of L.H., T.O. and S.O. was supported in part by the Director, 
Office of Science, Office of High Energy and Nuclear Physics, of the
U.S. Department of Energy under Contract DE-AC03-76SF00098, and in
part by the National Science Foundation under grant PHY-00-98840.
The work of S.O. was also supported by a National Science Foundation 
Graduate Research Fellowship.

\end{document}